\documentclass[manuscript]{aastex62}
\pdfoutput=1
        \usepackage{amsmath}
        \usepackage{amsfonts}
        \usepackage{amssymb}
        \usepackage{amsthm}
        \usepackage{latexsym}
        \usepackage{mathtools}

\graphicspath{{./}{figures/}}

\usepackage{xcolor}

\accepted{ApJ}

\shorttitle{NGC 4490 Double Nucleus}
\shortauthors{Lawrence et al.} 

\begin{document}

\title{Revealing the Double Nucleus of NGC 4490}

\correspondingauthor{Al Lawrence}
\email{allenl@iastate.edu}

\author[0000-0000-0000-0000]{Al Lawrence}
\affiliation{Iowa State University \\
Dept. of Physics \& Astronomy, 2323 Osborne Dr. \\
Ames, IA 50011-3160, USA}

\author[0000-0003-1539-3321]{C. R. Kerton}
\affiliation{Iowa State University \\
Dept. of Physics \& Astronomy, 2323 Osborne Dr. \\
Ames, IA 50011-3160, USA}

\author[0000-0000-0000-0000]{Curtis Struck}
\affiliation{Iowa State University \\
Dept. of Physics \& Astronomy, 2323 Osborne Dr. \\
Ames, IA 50011-3160, USA}

\author[0000-0000-0000-0000]{Beverly J. Smith}
\affiliation{East Tennessee State University \\
Dept. of Physics \& Astronomy, \\
Johnson City, TN 37614, USA}

\begin{abstract}
NGC 4490/85 (UGC 7651/48) or Arp 269 is well known for being one of the closest interacting/merging galactic systems. NGC 4490 has a  high star formation rate (SFR) and is surrounded by an enormous \ion{H}{1} feature stretching about 60 kpc north and south of the optically visible galaxies. Both the driver for the high SFR in NGC 4490 and the formation mechanism of the \ion{H}{1} structure are puzzling aspects of this system. We have used mid-infrared \emph{Spitzer} data to show that NGC 4490 has a double nucleus morphology. One nucleus is visible in the optical, while the other is only visible at infrared and radio wavelengths. We find the optical nucleus and the potential infrared visible nucleus have similar sizes, masses, and luminosities.   Both are comparable in mass and luminosity to other nuclei found in interacting galaxy pairs and much more massive and luminous compared with typical non-nuclear star-forming complexes. We examine possible origin scenarios for the infrared feature, and  conclude that it is  likely that NGC 4490 is itself a merger remnant, which is now interacting with NGC 4485. This earlier encounter provides both a possible driver for extended star formation in NGC 4490, and multiple pathways for the formation of the extended \ion{H}{1} plume.
\end{abstract}


\section{Introduction} \label{sec:int}

\begin{figure}
\plotone{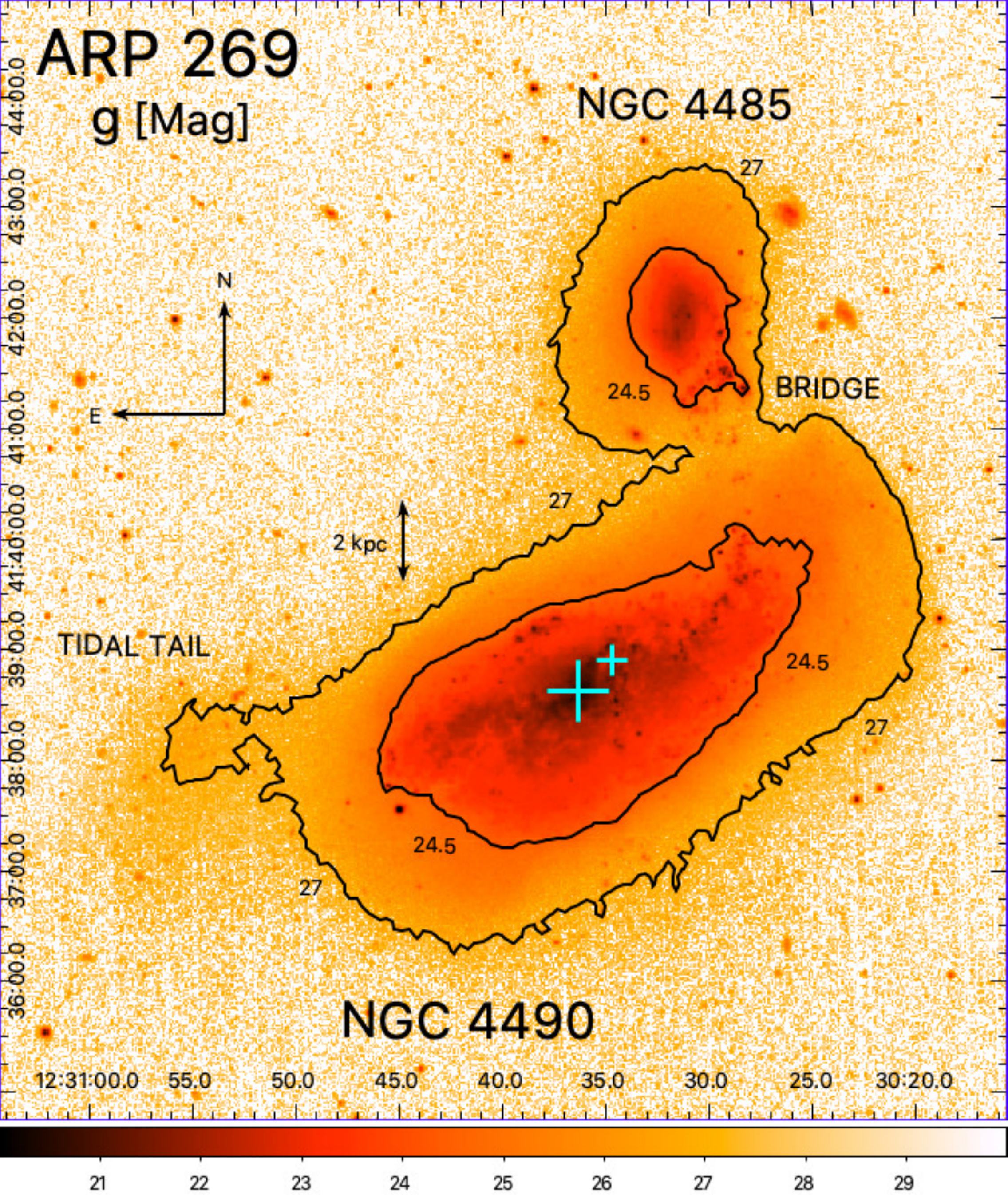}
\caption{SDSS $g$ mosaic image of Arp 269. The optical extent of the two galaxies is shown using contours at magnitude 24.5 and 27 (SDSS Pogson \citep{Pog1856} magnitude system). The large cyan cross marks the location of the optical nucleus (RA = 12$^{\mathrm h} 30^{\mathrm m} 36\fs2767$, DEC = $+41\degr 38\arcmin 37\farcs082$, J2000), and the small cyan cross (RA = 12$^{\mathrm h} 30^{\mathrm m} 35\fs6596$, DEC = $+41\degr 38\arcmin 54\farcs275$, J2000) shows the location of the dust-shrouded infrared nucleus. The 2 kpc line length is based on a distance of 9.2 Mpc. The mosaic was constructed using \textsc{montage} \citep{Ber2017}. \label{fig:ARP269}}
\end{figure}

NGC 4490/85  (UGC 7651/48) or Arp 269 \citep{Arp1966} is a relatively nearby ($d \approx 9.2$ Mpc; \citealt{Yim2016}) interacting galaxy system (see \autoref{tab:param} for a summary of all adopted physical and observational parameters).  As shown in \autoref{fig:ARP269}, it consists of a smaller irregular (IB 9) galaxy (NGC 4485) interacting with a larger barred spiral (SB 7) galaxy (NGC 4490, or colloquially the ``Cocoon Galaxy''). In this paper we present evidence that NGC 4490 is itself a late-stage merging system, and we discuss how this fact explains some puzzling aspects of the Arp 269 system.  

\begin{figure}
\plotone{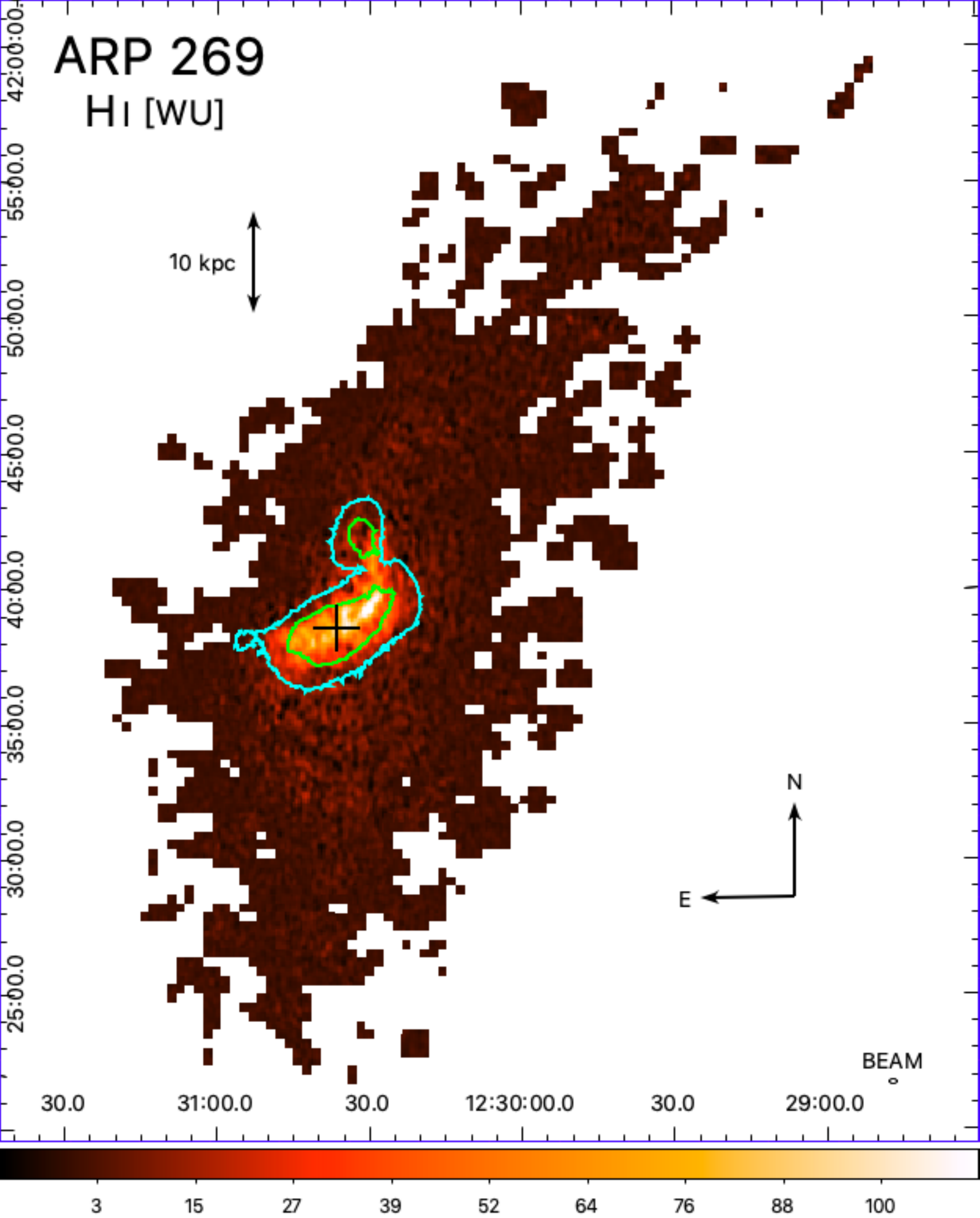}
\caption{Clipped and integrated (at a signal-to-noise level of 3)  \ion{H}{1} 21 cm emission around NGC 4490/85 from WHISP \citep{Van2001} in Westerbork Units (WU; 1 WU = 5 mJy beam$^{-1}$).
The total extent of the \ion{H}{1} plume is $\approx 120$ kpc for our adopted distance of 9.2 Mpc. The cyan and green contours represent the optical extent of NGC 4490/85 as shown in \autoref{fig:ARP269}, and the black cross marks the location of the NGC 4490 optical nucleus (see \autoref{fig:ARP269}). The beam, shown in the lower right, is $17.9 \times 12$ arcsec. Coordinate system is J2000. \label{fig:HIENVELOPE}}          
\end{figure}

One of the most striking features of NGC 4490/85, first noted by \citet{Huc1980} and \cite{Via1980}, is that it is embedded in an extensive envelope of neutral hydrogen (\ion{H}{1}). The \ion{H}{1} emission is elongated approximately perpendicular to the plane of NGC 4490 and has a total length of approximately 120 kpc (see \autoref{fig:HIENVELOPE}). One hypothesis for the origin of this structure is that it is the result of a star formation/supernova-driven outflow plume \citep{Cle1998}. An extensive multi-wavelength study of NGC 4490/85 by \cite{Cle1999} showed that the dynamic timescale for the formation of the plume ($\approx 6 - 7 \times 10^8$ yr) is consistent with a derived long-lived ($\approx 10^{8}$ yr) constant high rate of star formation (4.7 M$_{\odot}$ yr$ ^{-1}$), which we will call the NGC 4490 starburst hereafter. In addition, H$\alpha$ and X-ray observations (\citealt{Cle2002} and \citealt{Ric2010} respectively) show filaments and outflow features that are consistent with the outflow hypothesis.  Noting the minimal disruption of the optical disk of NGC 4490, \cite{Cle1999} argued that we are observing the first encounter between it and NGC 4485. They estimated that the time since the closest approach (i.e., pericentric passage) was $\approx 6 \times 10^7$ yr and estimated that the effects of the encounter on star formation in NGC 4490 would not become significant until $\approx 10^{8}$ yr after the closest approach. 

Alternatively the \ion{H}{1} emission may be from tidal features associated with the encounter between NGC 4485 and NGC 4490. \citet{Cle1998} briefly considered this idea, but simulations of galaxy collisions between unequal mass galaxies available at that time did not show elongated structures matching the \ion{H}{1} morphology.  This idea was followed up by \citet{Pea2018}, who modeled the interaction and found a plausible encounter history that results in a baryonic mass distribution similar to what is observed in \ion{H}{1}. In their best-match model there are two previous close encounters of the galaxies 1.3 Gyr and 0.19 Gyr ago, and the two galaxies are predicted to fully merge in another 0.4 Gyr. During the first passage a large tidal tail is formed primarily from NGC 4490 that eventually extends to a length of $\approx 200$ kpc. When viewed at a particular angle this tail is seen as a symmetric structure centered on NGC 4490 thus matching observations. One possible shortcoming of this model is the lack of an observed stellar component to the tidal feature as both stars and gas should be influenced by the tidal interaction; they suggest this apparent discrepancy could be the result of NGC 4490 having \textbf{a} gas-rich disk that is much larger than the stellar disk resulting in a gas dominated tidal feature. \citet{Pea2018} concluded that star-formation driven outflows may add material to the observed envelope, but they are not the source of the bulk of the material seen as the symmetric \ion{H}{1} structure. 

In this paper we use multi-wavelength images of NGC 4490 to show that the galaxy has two nuclei; one is visible primarily at optical wavelengths while the other one becomes prominent only at infrared (IR) and radio wavelengths due to extinction within NGC 4490. The use of IR imaging to investigate the structure of interacting galaxies dates back to the advent of IR imaging arrays. For example, \citet{Sta1990} used K-band images to show that there are two nuclei within the Arp 157 system, which led to it being used as the defining example of an intermediate-stage merger in the \citet{Hib1996} galactic merger sequence study. \citet{Sta1991} and \citet{Bus1992} conducted an IR imaging survey of approximately 180 Arp systems, but, since their survey was limited to objects that would fit within the $78 \times 84$ arcsec field of view of their detector, the relatively nearby Arp 269 system with its large angular extent was not examined. Radio continuum and H$\alpha$ emission associated with this second nucleus, which we will refer to as the `infrared' nucleus, have been observed before \citep{Sea1978,Duv1981}, but our study is the first to compare its physical properties to the optically-visible nucleus.  In the current paper, we show that it is comparable in mass and size, and we place this feature in the context of the history of the Arp 269 system. The fact that NGC 4490 could be a late-stage merging system, as suggested by the presence of two close nuclei, has implications for either picture of how the \ion{H}{1} plume formed. In the star-formation driven hypothesis, the merger could act as the driver for the necessary extended period of star formation. In the tidal debris picture the previous merger is a new part of the Arp 269 system that must be accounted for, and the resulting nucleus is a major structural feature of the system that must be matched in any simulation.

\begin{deluxetable*}{lll}
\tablecaption{NGC 4490/85 Basic Physical and Observational Parameters \label{tab:param}}
\tablewidth{0pt}
\tablehead{
\colhead{Parameter\tablenotemark{a}} & \colhead{Value} & \colhead{Reference}
}
\startdata
RA (J2000)                & 12$^{\mathrm h} 30^{\mathrm m} 36\fs277$ & See \autoref*{fig:ARP269} - optical nucleus  \\
DEC (J2000)               & $+41\degr 38\arcmin 37\farcs08$         &  See \autoref*{fig:ARP269} - optical nucleus \\
Hubble Class              &  SB(s)7p                                & \citet{Cle1998} \\
Hubble Class (4485)       &  IB(s)9p                                & \citet{Cle1998} \\
Inclination Angle         &  60\degr                                & \citet{Yim2016} \\
cz                        & 589 km s$^{-1}$                         & \citet{Fal1999} \\
Distance\tablenotemark{b} & 9.2~Mpc                                 & \citet{Yim2016} \\
D$_\mathrm{25}$           & 15 kpc                                  & \citet{Cle1998} \\
D$_\mathrm{25}$ (4485)    & 5.6 kpc                                 & \citet{Cle1998} \\ 
 \enddata
\tablenotetext{a}{For NGC 4490 unless otherwise indicated.}
\tablenotetext{b}{Derived using Hubble's Law and velocity field described in Appendix A of \citet{Mou2000}.}
\end{deluxetable*}

\section{Observations}  \label{sec:obs}

Unsurprisingly, given that Arp 269 is well known for being one of the closest interacting/merging galactic systems, there is an extensive collection of high-quality archival data available spanning a wide wavelength range. In \autoref{tab:FITS} we list the various data sets used in this study, and, for convenience, we briefly summarize the key characteristics of the data in the following paragraphs. Unless otherwise indicated, data were downloaded from the NASA/IPAC Extragalactic Database\footnote{http://ned.ipac.caltech.edu} (NED).

Ultraviolet (UV) data were obtained from the \textit{GALEX} mission. \textit{GALEX} was a NASA Small Explorer mission that operated between 2003 and 2012 making observations in two UV bands in the far-UV (FUV $\approx 154$ nm) and near-UV (NUV $ \approx $ 232 nm). The properties of the various \textit{GALEX} data products, including calibration details and resolution, are described in \cite{Mor2007}.  

Optical images of NGC 4490/85 (\textit{u, g, r, i,} and \textit{z}) were obtained from the Sloan Digital Sky Survey \citep[SDSS; ][]{Yor2000, Gun2006} SDSS-II Data Release 12 science archive server. In addition, we used narrow band H$\alpha$ observations of NGC 4490/85 from the Steward Observatory's 2.3-m Bok telescope at the Kitt Peak National Observatory \citep{Ken2008}.  

At infrared wavelengths we made use of data from the Two Micron All Sky Survey (2MASS) and the \emph{Spitzer} space telescope. 2MASS is an all-sky near-infrared survey in the J (1.25 $\mu$m), H (1.65 $\mu$m), and Ks (2.16 $\mu$m) bands \citep{Skr2006}.  Observations were conducted from two dedicated 1.3-m diameter telescopes located at Mount Hopkins, Arizona, and Cerro Tololo, Chile. The \textit{Spitzer} mission is described in \cite{Wer2004}. We made use of data from the 24, 70 and 160 $\mu$m bands of the Multiband Imaging Photometer \citep[MIPS;][]{Rie2004}, and data from the 3.6, 4.5, 5.6, and 8.0 $\mu$m bands of the Infrared Array Camera \citep[IRAC;][]{Faz2004}. The 3.6 and 4.5 $\mu$m data are described in \citet{She2010}, and the observations in other bands are described in \citet{Dal2016}.

At radio wavelengths, the 8.46 GHz continuum observation from the Very Large Array (VLA) was obtained from the NRAO VLA Archive Site. \ion{H}{1} integrated 21-cm emission data (clipped at a signal to noise level of 3) was obtained from the WHISP (Westerbork observations of neutral Hydrogen (\ion{H}{1}) in Irregular and SPiral galaxies) survey \citep{Van2001} using the Westerbork Synthesis Radio Telescope (WSRT). Surface brightness in the WHISP maps are in Westerbork Units (WU), where 1 WU = 5 mJy beam$^{-1}$ \citep{Yim2016}. Images were created using a beam size of $17.9 \times 12$ arcsec (ellipse major $\times$ minor axes), and the data cube channel spacing is 4.14 km s$^{-1}$. Details of the WHISP observations and data reduction can be found in \cite{Van2001}, \cite{Swa2002}, and \cite{Noo2005}. 

\begin{figure*}
\gridline{\fig{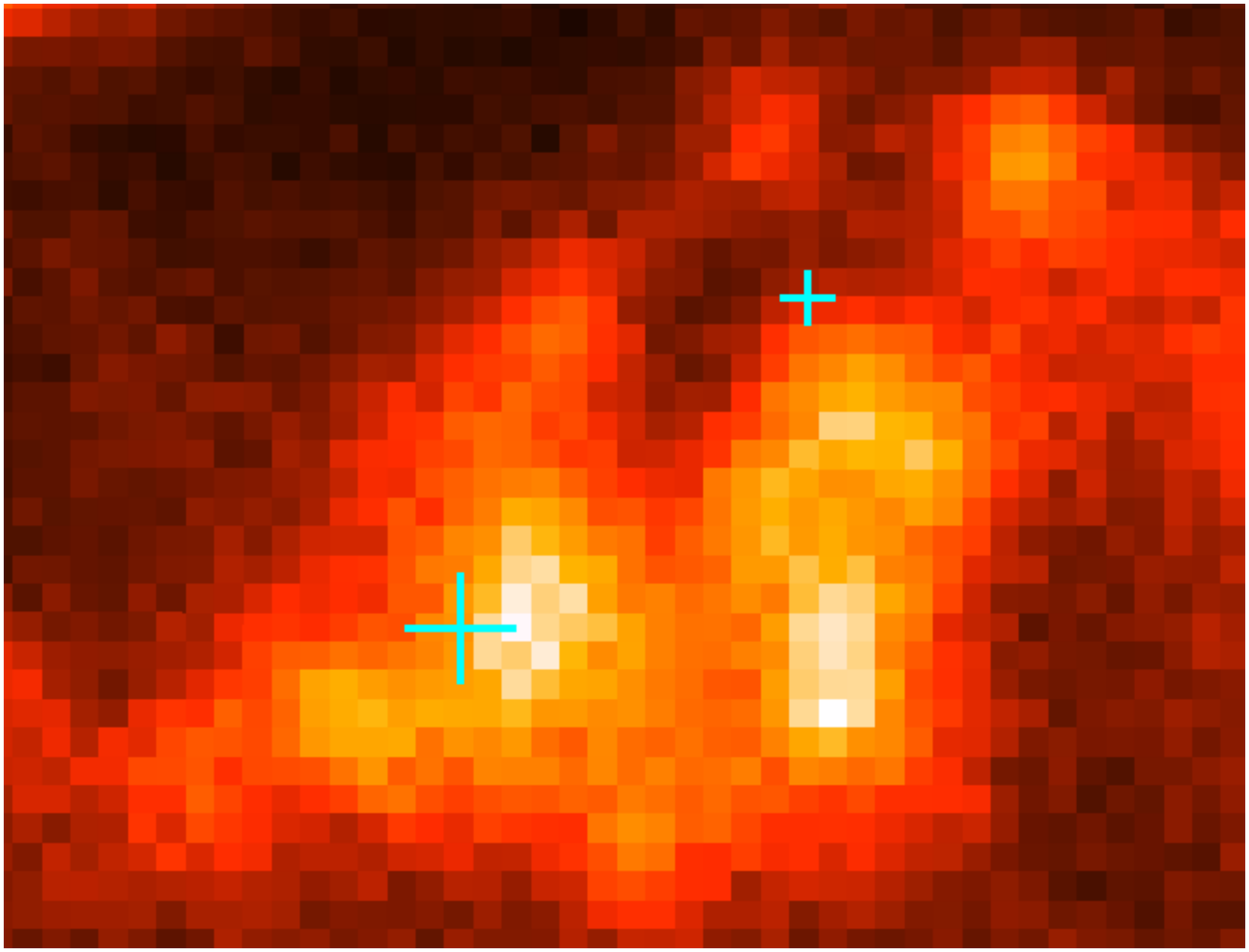}{0.29\textwidth}{FUV}
          \fig{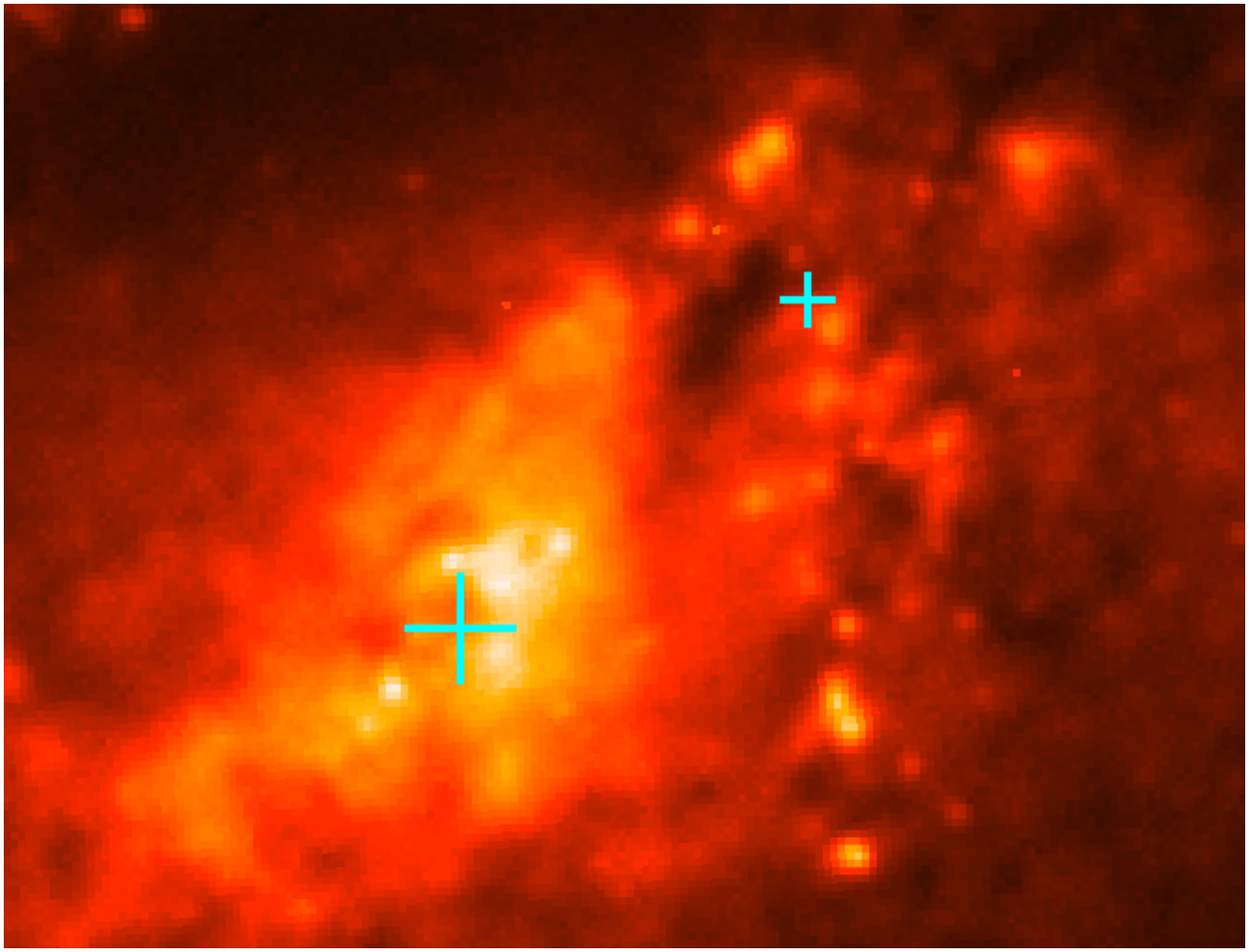}{0.29\textwidth}{g}
          \fig{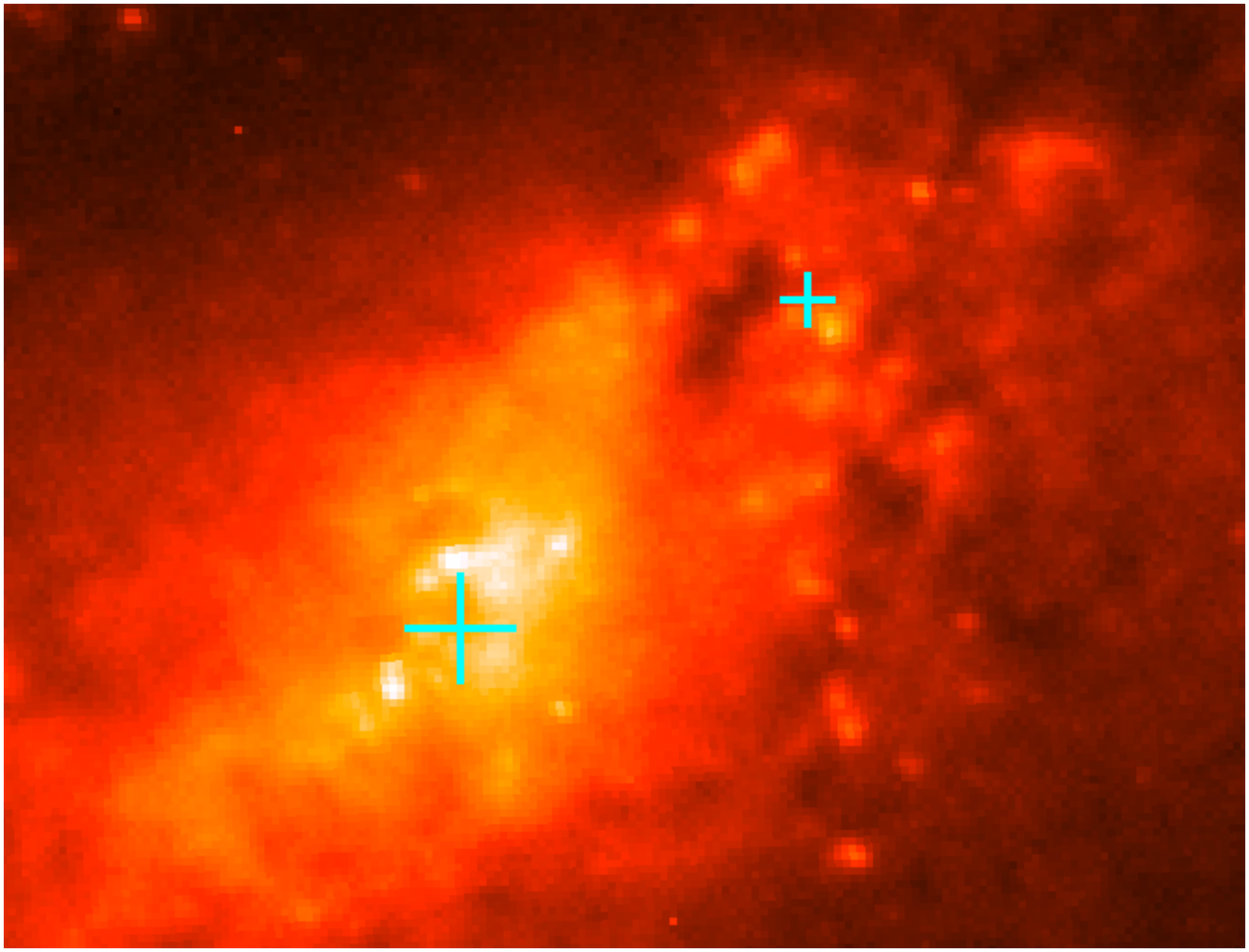}{0.29\textwidth}{i}
          }
\gridline{\fig{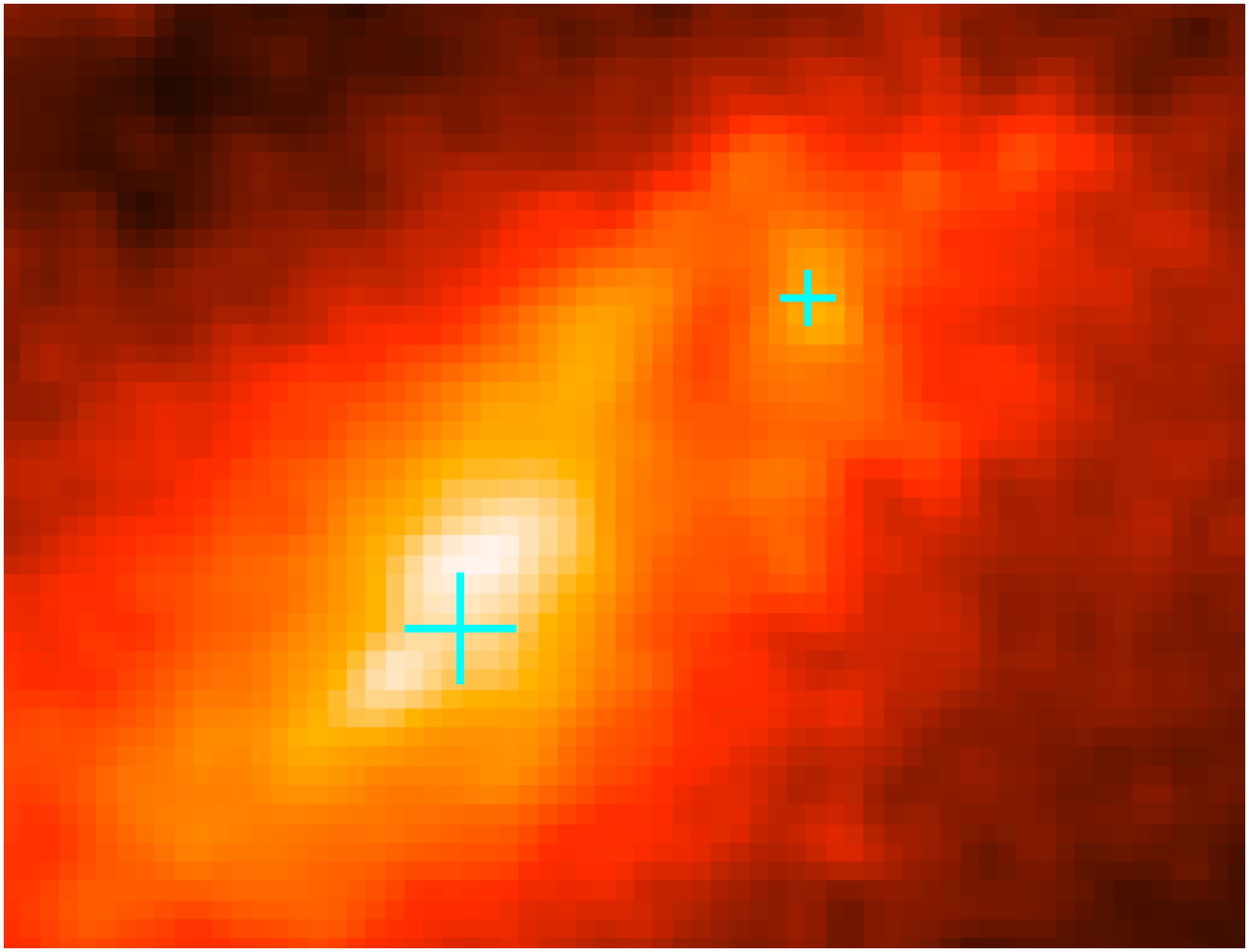}{0.29\textwidth}{J}
          \fig{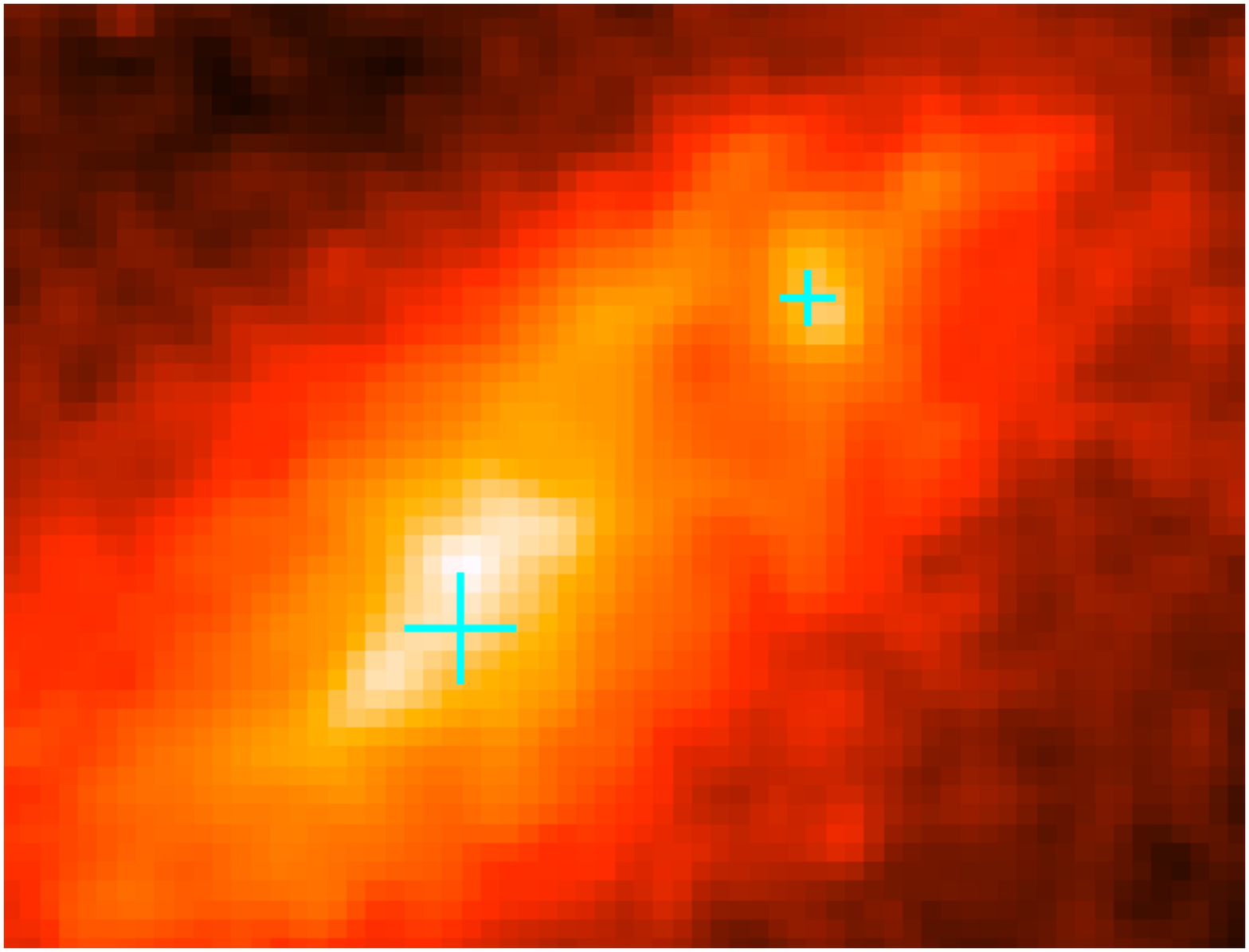}{0.29\textwidth}{H}
          \fig{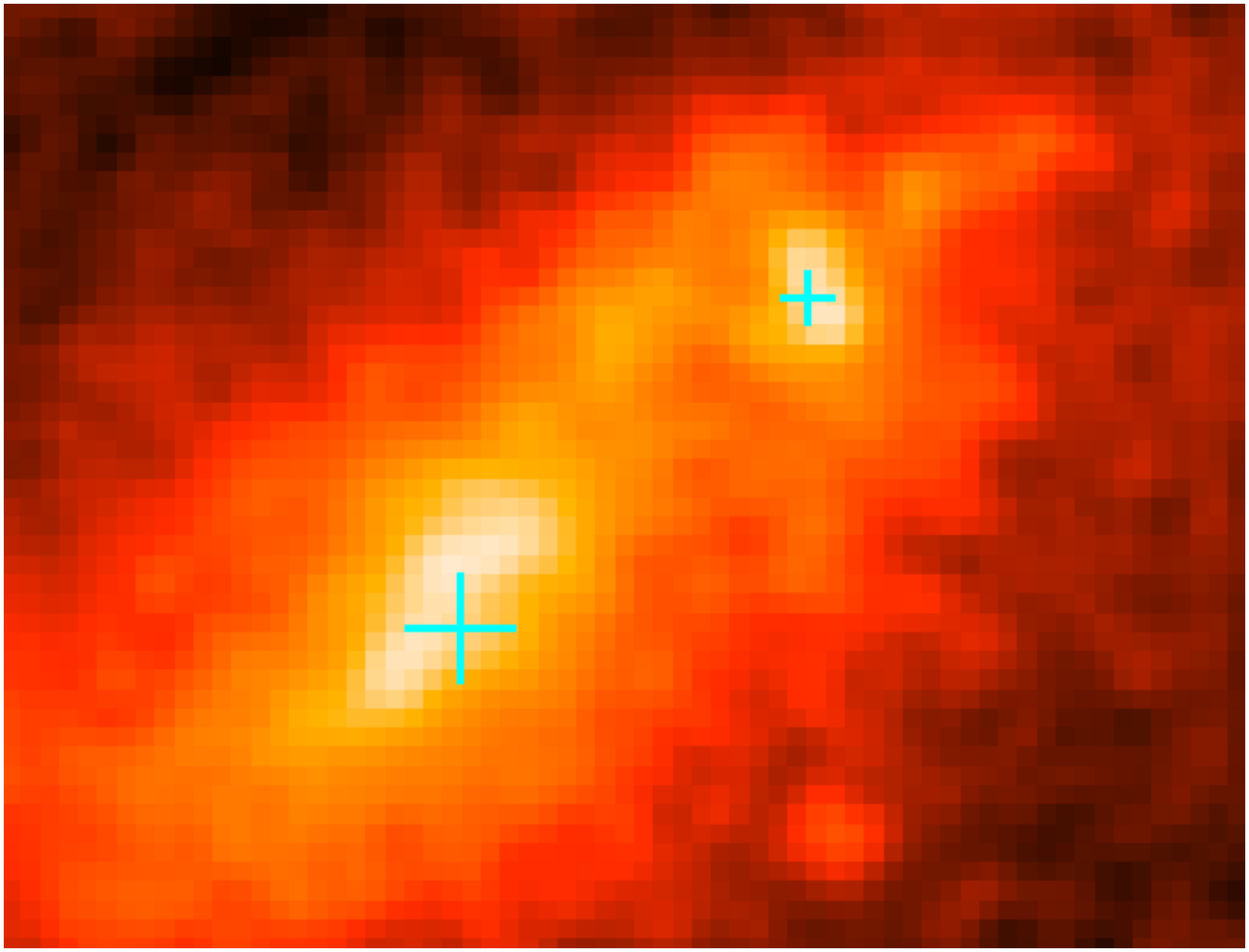}{0.29\textwidth}{Ks}
          }
\gridline{\fig{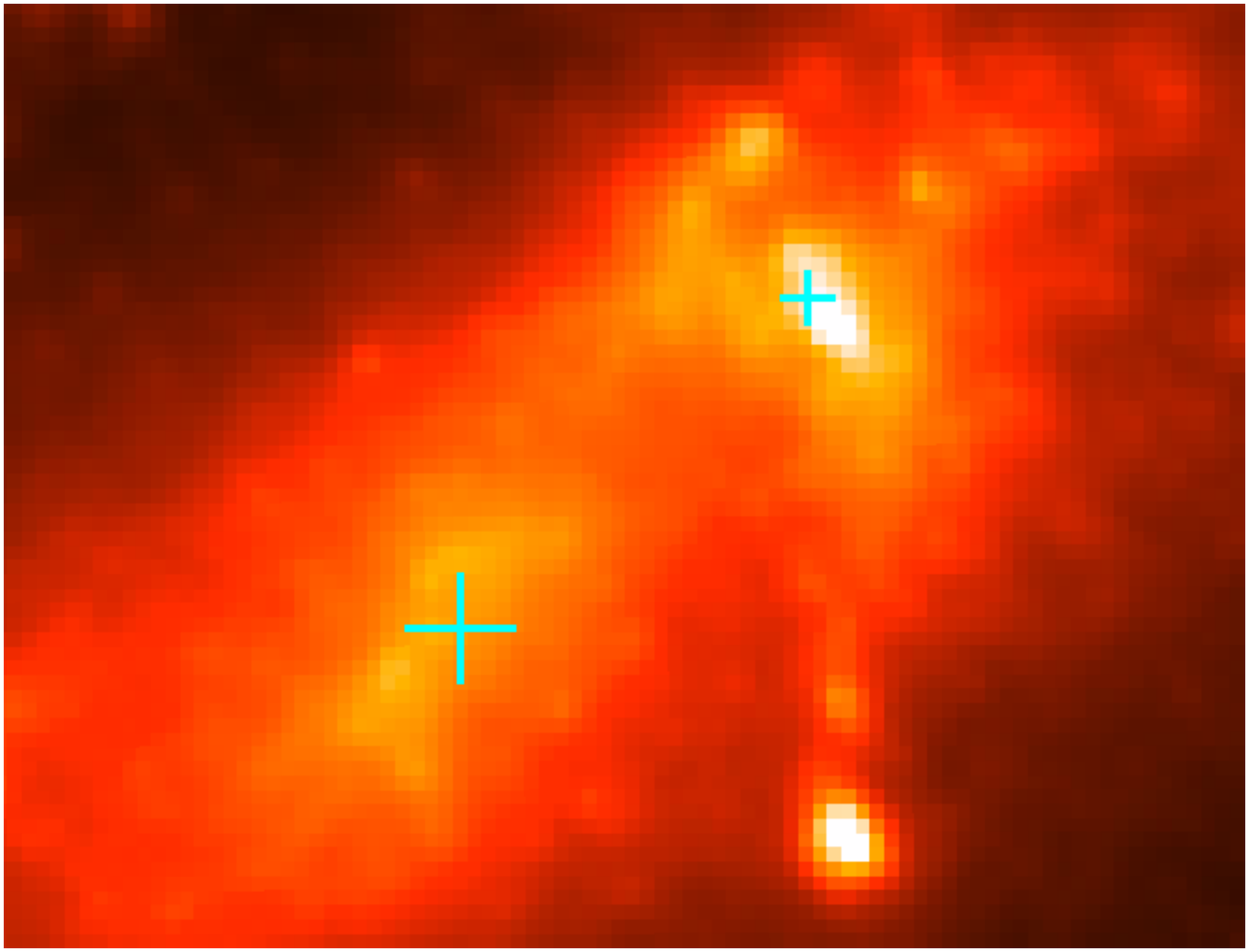}{0.29\textwidth}{3.6 $ \mu $m}
          \fig{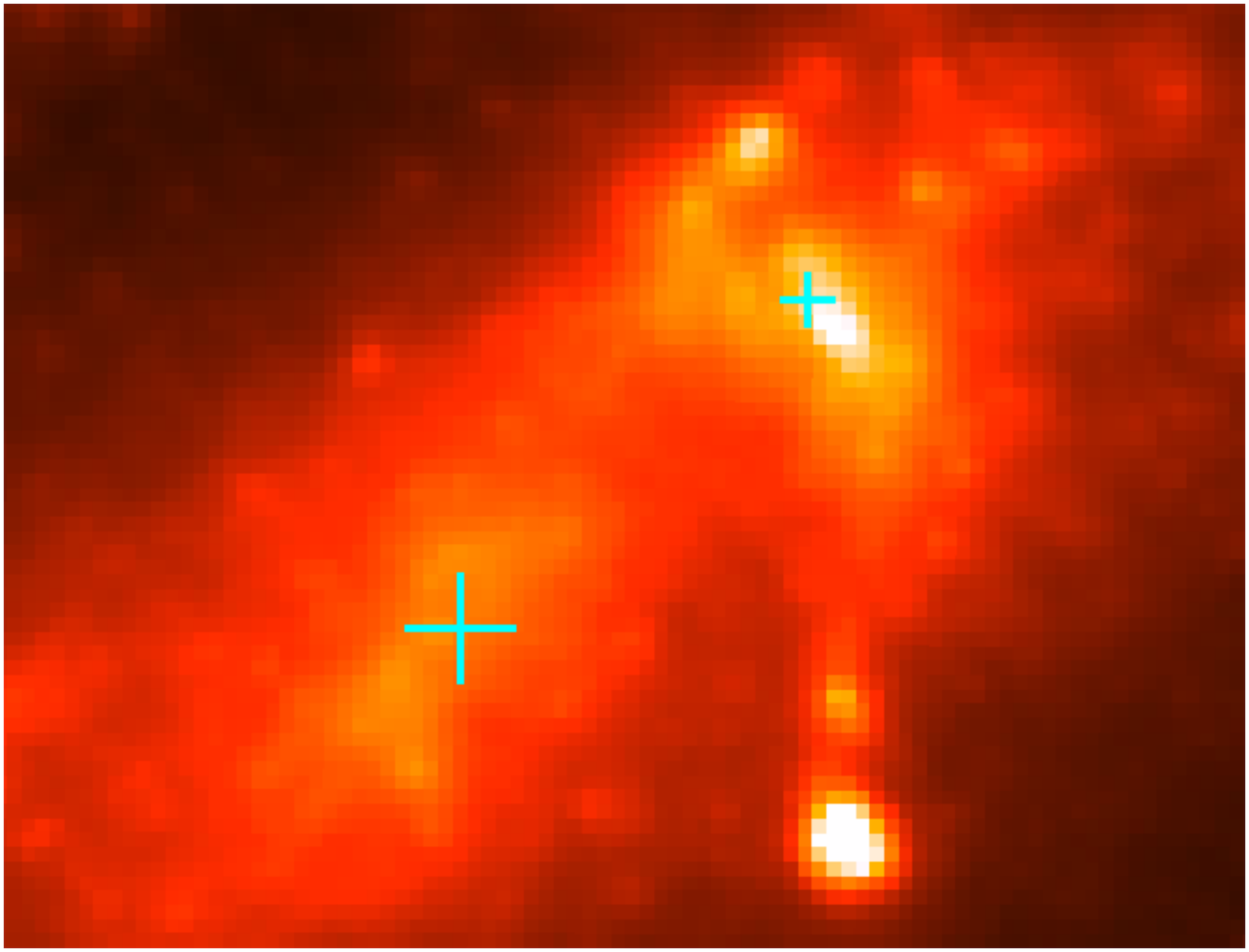}{0.29\textwidth}{4.5 $ \mu $m}
          \fig{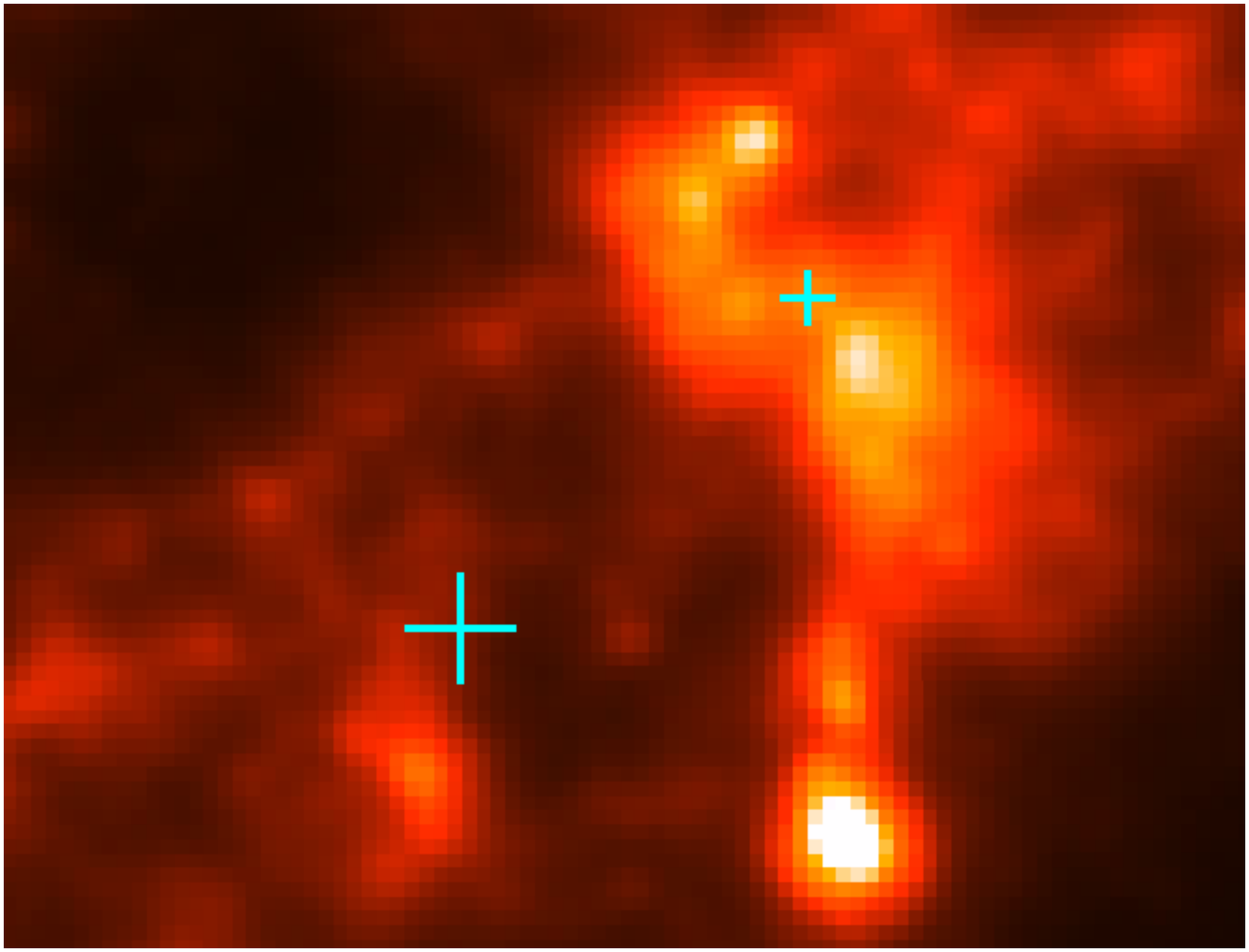}{0.29\textwidth}{8.0 $ \mu $m}
          }
\gridline{\fig{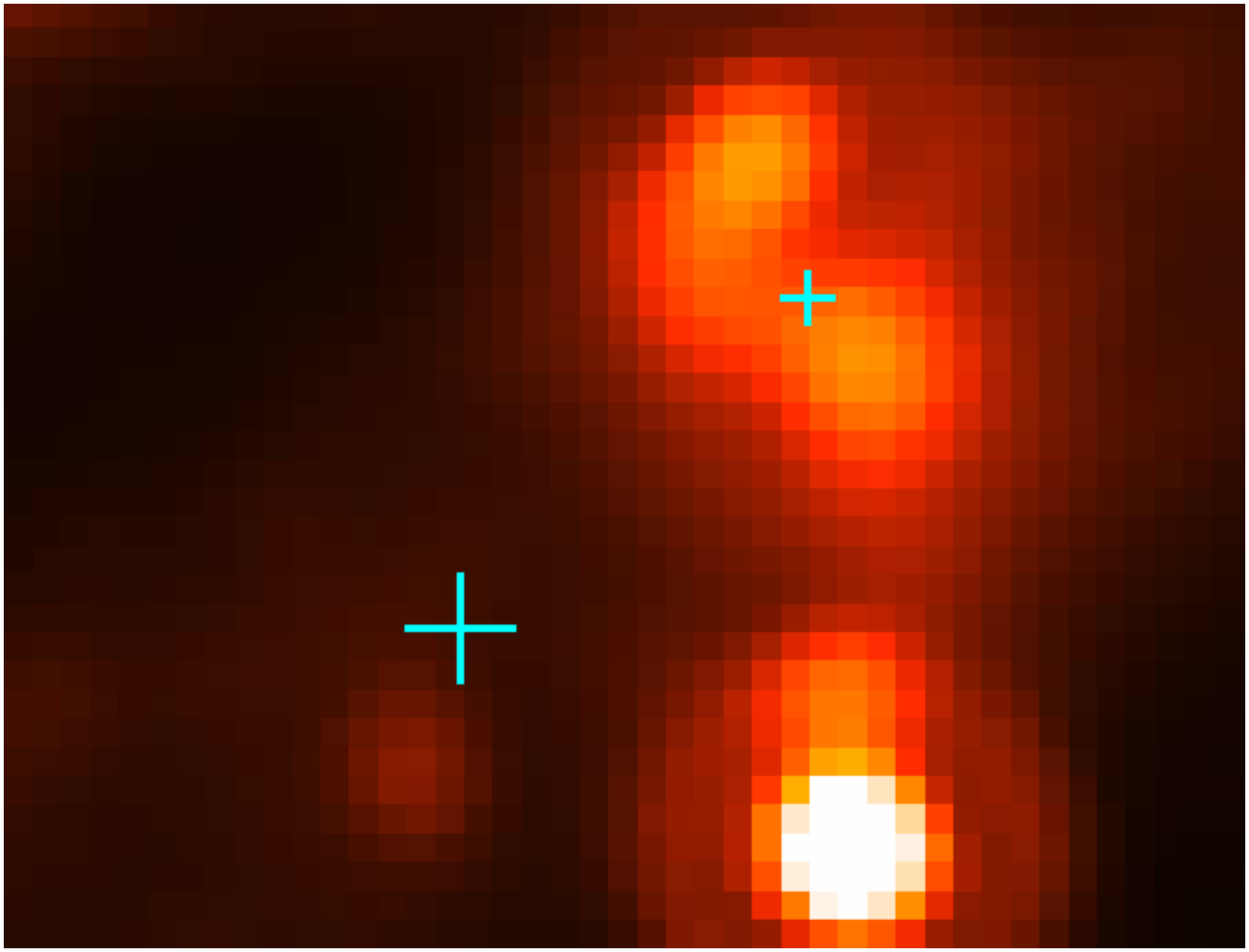}{0.29\textwidth}{24 $ \mu $m}
          \fig{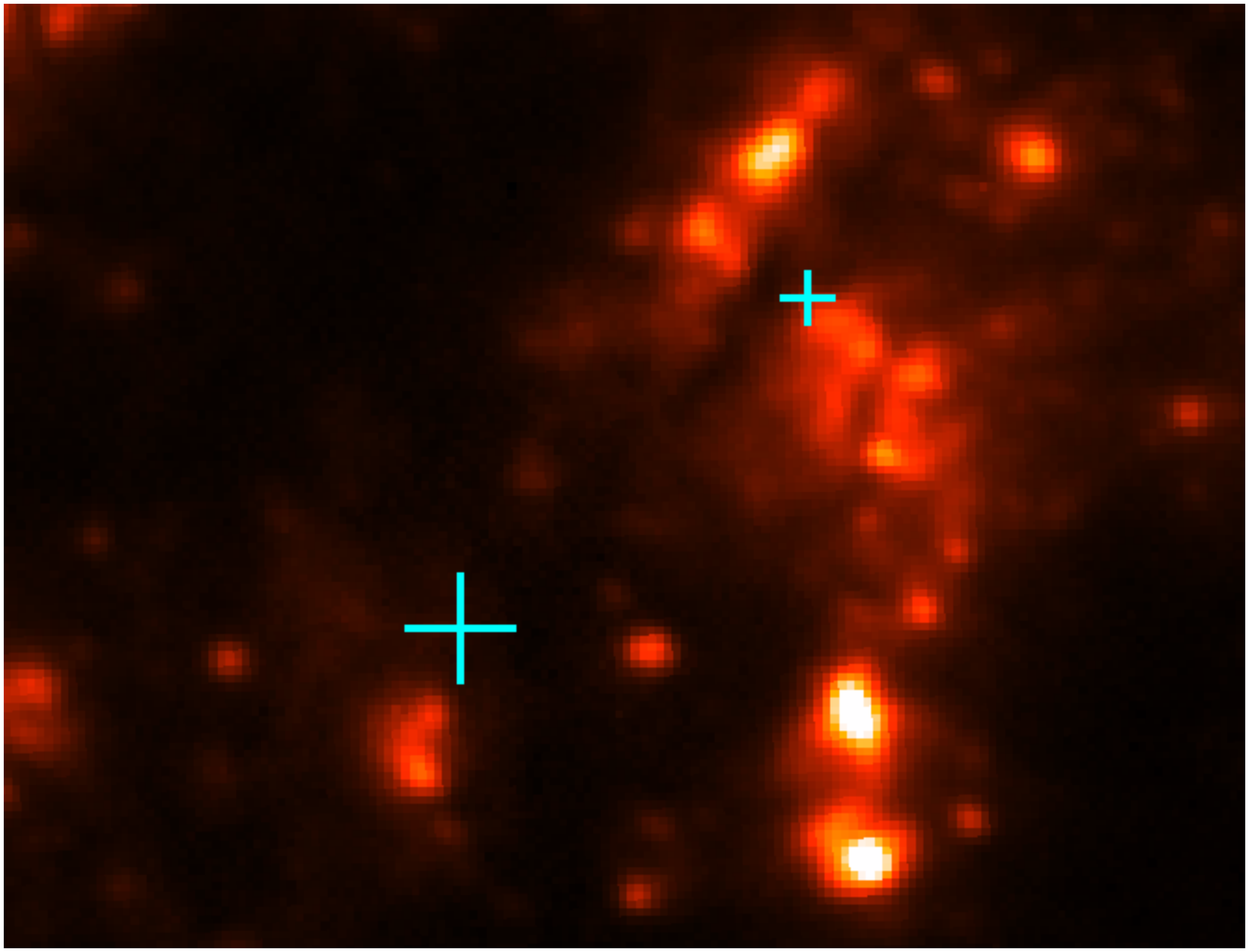}{0.29\textwidth}{H$ \alpha $}
          \fig{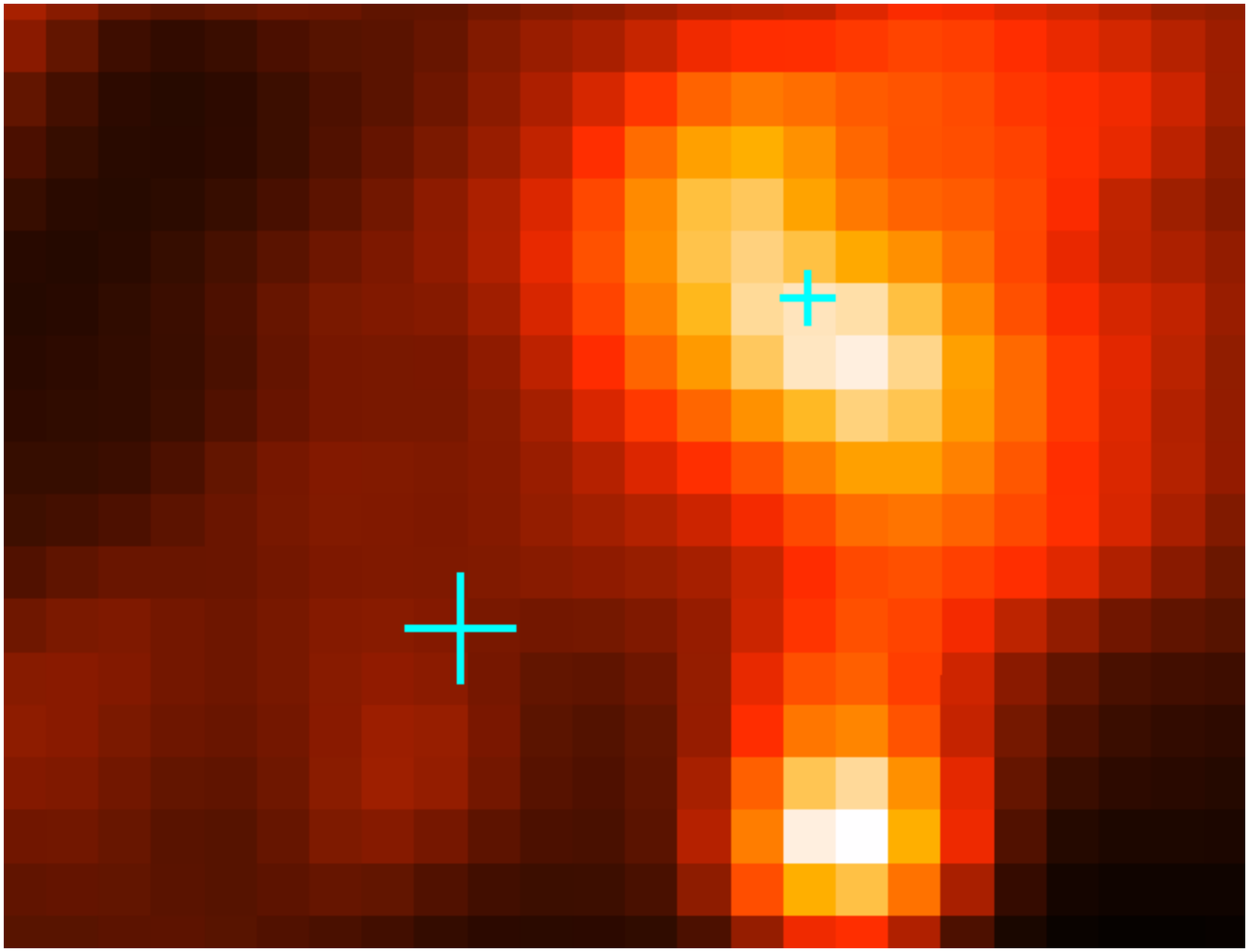}{0.29\textwidth}{8.46 GHz}
          }
\caption{NGC 4490 multi-wavelength double nucleus images. Each panel is $64.9 \times 49.3$ arcsec in size ($\approx 2.9 \times 2.2$ kpc at 9.2 Mpc) centered at RA=12$^{\mathrm h} 30^{\mathrm m}35\fs478$, DEC = $+41\degr38\arcmin45\farcs22$ (J2000).  The large and small cyan crosses mark the locations of the optical and infrared nucleus respectively (see \autoref{fig:ARP269}). \label{fig:Double Nucleus}}
\end{figure*}

\autoref{fig:Double Nucleus} illustrates how the appearance of the central region of NGC 4490 changes dramatically with wavelength. The optical and the infrared nuclei are both best seen in the near-IR  images. At longer wavelengths emission becomes dominated by heated dust in the interstellar medium and by emission associated with individual \ion{H}{2} regions, which are evident in the H$\alpha$ image.  At shorter wavelengths dust obscures the infrared nucleus and the galaxy looks more like a normal galaxy with a single dominant nucleus. 

\begin{deluxetable*}{lcclc}
\tablecaption{NGC 4490/85 Data Sets \label{tab:FITS}}
\tablewidth{0pt}
\tablehead{
\colhead{Source} & \colhead{Band} & \colhead{$ \lambda $} & \colhead{FITS Data Set} & \colhead{Resolution} }
\startdata
\textit{GALEX} & FUV & 1539 $ \AA $  & NGC$\textunderscore$4490$\textunderscore$I$\textunderscore$FUV$\textunderscore$d2009.fits & 4.2\arcsec\\
     & NUV & 2316 $ \AA $  & NGC$\textunderscore$4490$\textunderscore$I$\textunderscore$NUV$\textunderscore$d2009.fits&5.3\arcsec\\
SDSS & $ u $ &  3560 $ \AA $  & frame-u-003840-2-0188.fits& $\approx1.5\arcsec$ \\
     & $ g $ &  4680 $  \AA $  &  frame-g-003840-1-0188.fits\tablenotemark{a}&$ \approx1.5\arcsec$ \\
     &  &  $  $  &  frame-g-003840-1-0189.fits\tablenotemark{a}\\
     &  &  $  $  & frame-g-003840-2-0188.fits\tablenotemark{a}\tablenotemark{b}\\
     &  &  $  $  & frame-g-003840-2-0189.fits\tablenotemark{a}\\
     & $ r $ &  6180 $ \AA $  & frame-r-003840-2-0188.fits&$\approx 1.5\arcsec$\\
Bok & H$ \alpha $ & 6580 $  \AA $  & NGC$\textunderscore$4490$\textunderscore$I$\textunderscore$Ha$\textunderscore$d2009.fits&$\approx 1.5\arcsec$\\    
SDSS & i & 7500 $  \AA $  & frame-i-003840-2-0188.fits&$ \approx 1.5\arcsec $\\
     & $ z $ &  8790 $  \AA $  &  frame-z-003840-2-0188.fits&$ \approx 1.5\arcsec$\\
2MASS& J & 1.25 $ \mu $m  &2MASS$\textunderscore$NGC$\textunderscore$4490$\textunderscore$J.fits&$ \approx 2.5\arcsec$\\
     & H & 1.65 $ \mu  $m  & 2MASS$\textunderscore$NGC$\textunderscore$4490$\textunderscore$H.fits&$ \approx 2.5\arcsec$ \\
     & Ks & 2.16 $ \mu  $m  & 2MASS$\textunderscore$NGC$\textunderscore$4490$\textunderscore$K.fits&$ \approx 2.5\arcsec$\\  
\textit{Spitzer} & IRAC & 3.6 $ \mu $m & NGC$\textunderscore$4485$\textunderscore$I$\textunderscore$IRAC$\textunderscore$3.6$\textunderscore$srh2010.fits&1.66\arcsec\\
     &  & 4.5 $ \mu $m &NGC$\textunderscore$4485$\textunderscore$I$\textunderscore$IRAC$\textunderscore$4.5$\textunderscore$srh2010.fits&1.72\arcsec\\
     &  & 5.8 $ \mu $m & NGC$\textunderscore$4485$\textunderscore$I$\textunderscore$IRAC$\textunderscore$5.8$\textunderscore$d2009.fits&1.88\arcsec\\
     &  & 8.0 $ \mu $m & NGC$\textunderscore$4485$\textunderscore$I$\textunderscore$IRAC$\textunderscore$8.0$\textunderscore$d2009.fits&1.98\arcsec\\
     &  MIPS & 24 $ \mu $m & NGC$\textunderscore$4490$\textunderscore$I$\textunderscore$MIPS24$\textunderscore$d2009.fits&$ \approx 6\arcsec$\\ 
      &   & 70 $ \mu $m & NGC$\textunderscore$4490$\textunderscore$I$\textunderscore$MIPS70$\textunderscore$d2009.fits&$ \approx 18\arcsec$\\ 
     &   & 160 $ \mu $m & NGC$\textunderscore$4490$\textunderscore$I$\textunderscore$MIPS160$\textunderscore$d2009.fits&$ \approx 40\arcsec$\\
   VLA  & 8.46 GHz  & 3.54 cm &8.46I7.98$\textunderscore$AA0181$\textunderscore$1995MAR16$\textunderscore$1$\textunderscore$79.4U2.60M.imfits&$\approx 8 \arcsec$\\            
WHISP & \ion{H}{1}  & 21.147 cm & UGC$\textunderscore$07651$\textunderscore$FR$\textunderscore$I$\textunderscore$21cm$\textunderscore$h2002.fits & $\approx 15\arcsec$ \\           
\enddata
\tablenotetext{a}{Used for \autoref*{fig:ARP269} mosaic.}
\tablenotetext{b}{Used for \autoref*{fig:Double Nucleus}.}
\end{deluxetable*}

\section{Analysis} \label{sec:ana}

\begin{figure}
    \plotone{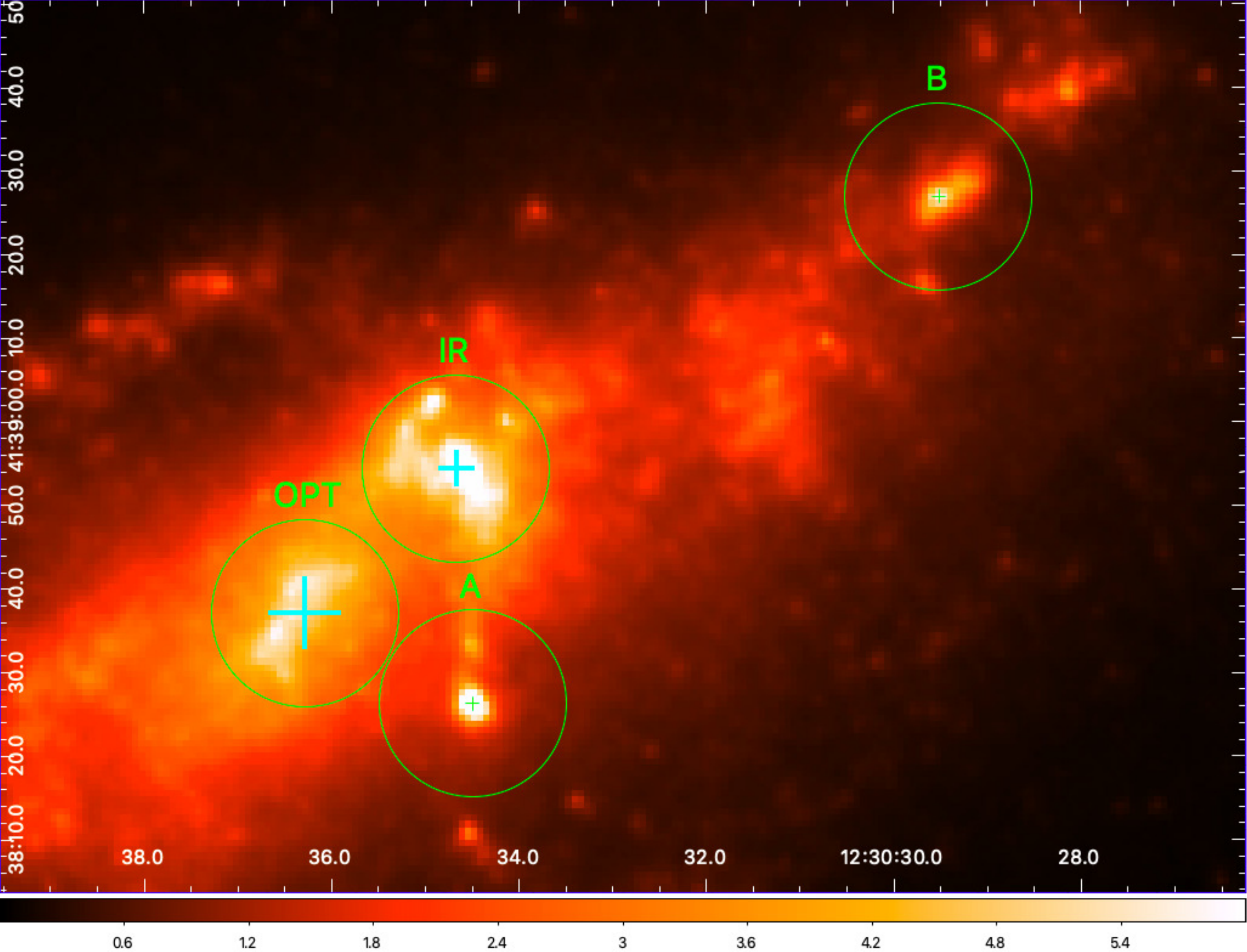}
    \caption{NGC 4490 3.6~$\mu$m image ($ \approx $ 6.7 x 4.8 kpc at 9.2 Mpc) showing regions of interest. Apertures (500 pc radius) are shown surrounding the optical nucleus (OPT), infrared nucleus (IR), and the giant \ion{H}{2} regions A  and B.  The large and small cyan crosses mark the locations of the optical and infrared nucleus respectively (see \autoref{fig:ARP269}).  The units for the image are MJy sr$^{-1}$. Coordinate system is J2000.}
    \label{fig:aps}
\end{figure}

\begin{figure}
\plotone{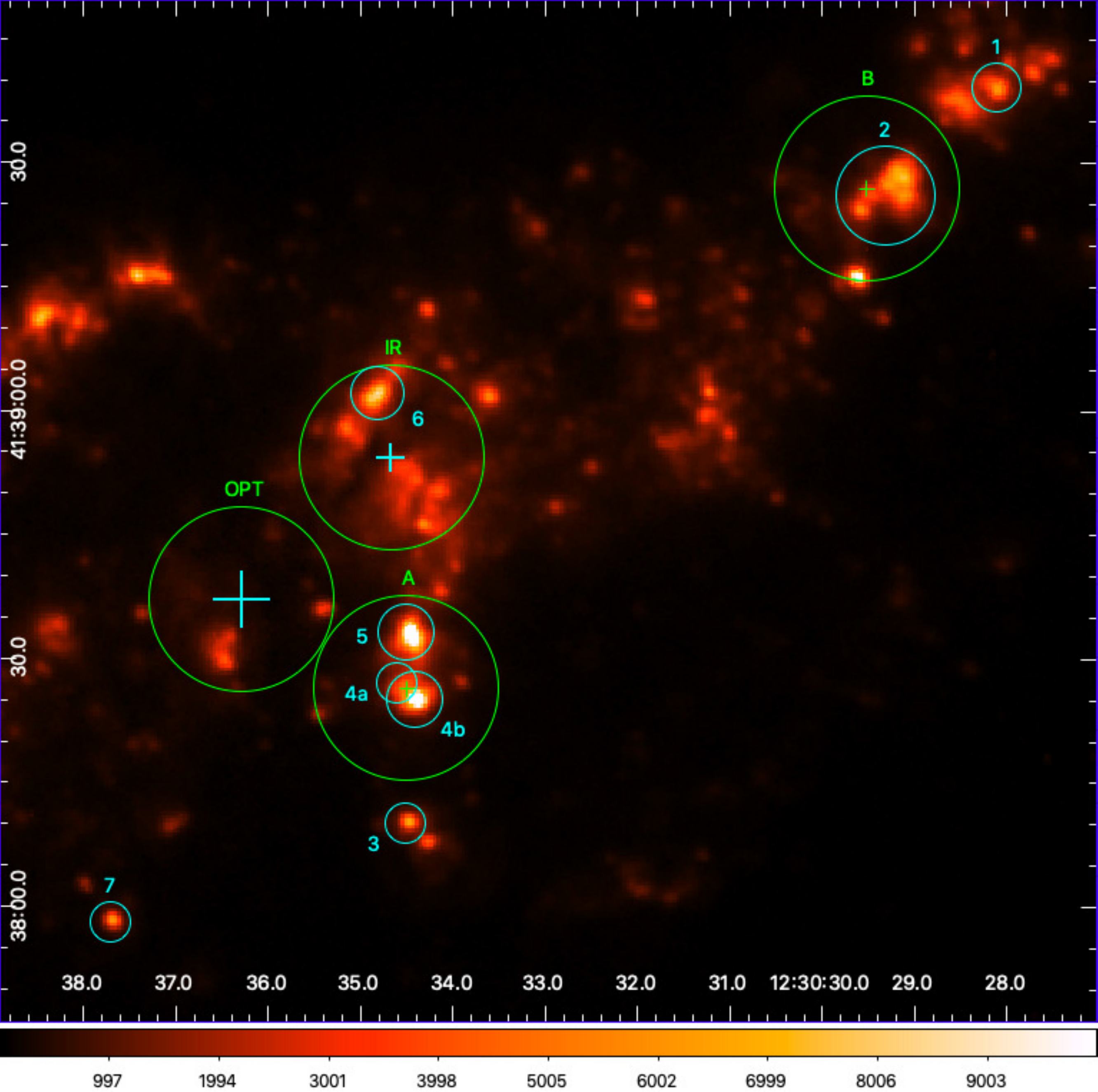}
\caption{NGC 4490 H$ \alpha $ image  ($ \approx $ 5.9 x 5.5 kpc at 9.2 Mpc) showing \ion{H}{2} regions (numbered; cyan circles) from \cite{Cle2002} and 500 pc radius apertures (green circles) from \autoref{fig:aps}. \ion{H}{2} regions sizes are approximate.  Units for the H$ \alpha $ image colorbar are in  counts $(1000 \mathrm{s})^{-1} $,  where 1 count s$^{-1} = 6.252 \times 10^{-16}$ erg s$^{-1}$ cm$^{-2}$ \AA$^{-1}$.   The large  and small cyan crosses mark the locations of the optical and infrared nucleus respectively (see \autoref{fig:ARP269}).  Coordinate system is J2000.  \label{fig:HASOURCE}}          
\end{figure}

Since the optical and infrared nuclei are equally prominent at 3.6 $\mu$m we used the \emph{Spitzer} image in that band to define the apertures used for our analysis. \autoref{fig:aps} shows the central and northwest portions of NGC 4490 at 3.6 $\mu$m. We use 500 pc radius apertures centered on the optical nucleus (OPT) and the infrared nucleus (IR). These are shown along with similar apertures around two more compact regions (A and B). These two compact regions correspond to regions 4a+4b+5 and region 2 from the radio study of \citet{Cle2002} respectively. An H$\alpha$ image of NGC 4490 in \autoref{fig:HASOURCE} shows the position and approximate extent of the  \citet{Cle2002} \ion{H}{2} regions along with our apertures. We find the H$\alpha$ luminosity ($\textit{L}_{\text{H}\alpha}$), calculated using the apertures shown in \autoref{fig:HASOURCE}, is $1.43\times10^{40}$ and $1.23\times10^{40}$ erg~s$^{-1}$ for regions A and B respectively. These $L_{\text{H}\alpha}$ values, and the compact nature of each source, are consistent with these regions being giant \ion{H}{2} regions (e.g., see Table 9.1 in \citealt{Con2008}). The $L_{\text{H}\alpha}$ of the \ion{H}{2} regions A and B  are both similar to the $\textit{L}_{\text{H}\alpha}$ of 30 Doradus (1.5 x 10$ ^{40} $ erg s$^{-1}$), which was measured in a 370 pc diameter region \citep{Con2008, Ken1984}.    The $L_{\text{H}\alpha}$ values calculated for the 500 pc regions (OPT, IR, A and B) and the galaxies (NGC 4490 and NGC 4485) are shown in column 22 of  \autoref{tab:SEDMASSLUM}.   The combined $\textit{L}_{\text{H}\alpha}$ of the four 500 pc aperture regions (IR, OPT, A, and B) accounts for 25\% of the total $L_{\text{H}\alpha}$ of NGC 4490. 

\subsection{Stellar Mass} \label{sec:STARMASS}

The stellar mass ($\mathrm{M}_\star$) of the various regions can be estimated from \emph{Spitzer} 3.6 and 4.5 $\mu$m flux densities ($F_{3.6}$ and $F_{4.5}$) using the \citet{Esk2012} calibration, which was based on resolved observations of the Large Magellanic Cloud (LMC), 
\begin{equation}
    \label{equ:STARMASS}
     \mathrm{M}_\star= 10^{5.65} F_{3.6}^{2.85} F_{4.5}^{-1.85} \left(\frac{D}{0.05}\right)^2 \;\; \mathrm{M_\sun}
\end{equation}
where the flux densities are in Jy, $D$ is the distance to NGC 4490 in Mpc, and the adopted distance to the LMC is 50 kpc. Photometry was obtained from sky-subtracted images for each of the nuclei at 3.6 and 4.5 $\mu$m using the 500 pc apertures shown in \autoref{fig:aps} with no local background subtraction. This means that the fluxes we measure include light from the inner portion of the undefined disk, the inner bulge (if any), in addition to the nucleus itself.  In the following analysis, we use the term `nucleus' to refer generically to this part of the galaxy.  We also measured $F_{3.6}$ and $F_{4.5}$ for NGC 4490 and NGC 4485 using elliptical apertures of $9.12 \times 4.31 $ kpc and $3.07 \times 2.42$ kpc respectively (semi-major $\times$ semi-minor axis). In addition to the $\approx 1$ Mpc uncertainty in the distance to NGC 4490 (e.g., \citet{Cle1998} uses 8~Mpc), the main source of error in the mass estimate is a $1\sigma$ uncertainty of up to $\approx 30\%$ caused by intrinsic scatter in the calibration \citep{Esk2012}. All of the flux density measurements and derived masses are summarized in  \autoref{tab:SEDMASSLUM}.

We find that the mass of the optical and infrared nuclei are comparable ($\mathrm{M}_\mathrm{OPT} \approx 1.1 \mathrm{M}_\mathrm{IR}$), and each nucleus contains about 10 percent of the stellar mass of NGC 4490. The infrared and optical nuclei are each also 3 -- 5 times more massive than either of the two \ion{H}{2} regions, which each contain only 2 -- 3 percent of the stellar mass of NGC 4490. As expected, the total stellar mass of NGC 4490 is much larger ($\approx 15$ times) than NGC 4485, but it is approximately 5 times smaller than the Milky Way's total stellar mass of $6.08\pm1.14 \times 10^{10} \ \mathrm{M} _{\odot}$ \citep{Lic2015}. Both NGC 4490 and NGC 4485 have stellar masses comparable ($\approx 5$ and $\approx 1.5$ times larger) to the Large and Small Magellanic Clouds respectively \citep{Van2002,Sta2004,Pea2018}.

\begin{splitdeluxetable*}{lcccBlll  lllllllBllllll lcc}
\tablewidth{0pt} 
\tablecaption{SED Photometry, Stellar Mass, \& H$\alpha$ Luminosity \label{tab:SEDMASSLUM}}
\tablehead{\colhead{} & Aperture & RA (Center) & DEC (Center) & \colhead{FUV} & \colhead{NUV} & \colhead{u} & \colhead{g} & \colhead{r} & \colhead{i} & \colhead{z} & \colhead{J}& \colhead{H} & \colhead{Ks} & \colhead{3.6 $\mu $m} & \colhead{4.5 $\mu$m} & \colhead{5.8 $\mu$m} & \colhead{8 $\mu$m} & \colhead{24 $\mu$m} & \colhead{70 $\mu$m} & \colhead{160 $\mu$m} & \colhead{$\textit{L}_{\text{H}\alpha}$}& \colhead{$\log(\mathrm{M_\star}/\mathrm{M_\sun})$}  \\
\colhead{Object} & \colhead{Radius (at 9.2 Mpc)} & \colhead{J2000} & \colhead{J2000} & \colhead{(mJy)} & \colhead{(mJy)} & \colhead{(mJy)} & \colhead{(mJy)} & \colhead{(mJy)} & \colhead{(mJy)} & \colhead{(mJy)} & \colhead{(mJy)} & \colhead{(mJy)} & \colhead{(mJy)} & \colhead{(mJy)} & \colhead{(mJy)} & \colhead{(mJy)} & \colhead{(mJy)} & \colhead{(mJy)} & \colhead{(mJy)} & \colhead{(mJy)} & \colhead{(erg s$^{-1}$)} } 
\startdata 
Optical Nucleus (OPT)                &  500 pc                         & 12$^{\mathrm h} 30^{\mathrm m}36\fs2767$ & $+41\degr38\arcmin37\farcs082$ & 2.64  & 2.30 & 13.3 & 35.1 & 45.7 & 53.9 & 62.2 & 68.9 & 77.1 & 61.6 & 35.4 & 23.9 & 39.5 & 75.8 & 118  & 1860  & 1737  & 6.39E+39  & 9.04 \\
Infrared Nucleus (IR)                &  500 pc                         & 12$^{\mathrm h} 30^{\mathrm m}34\fs6596$ & $+41\degr38\arcmin54\farcs275$ & 1.69  & 3.02 & 7.72 & 16.4 & 24.2 & 28.3 & 35.8 & 45.4 & 55.6 & 49.9 & 38.0 & 27.9 & 80.7 & 198  & 471  & 4045  & 2571  & 1.95E+40  & 9.01 \\
H\footnotesize{II} Region A &  500 pc                         & 12$^{\mathrm h} 30^{\mathrm m}34\fs4903$ & $+41\degr38\arcmin26\farcs287$ & 1.58  & 1.26 & 5.58 & 11.6 & 15.2 & 16.4 & 18.7 & 20.8 & 22.6 & 19.8 & 16.1 & 13.5 & 39.9 & 104  & 505  & 2059  & 1772  & 1.43E+40  & 8.53 \\
H\footnotesize{II} Region B & 500 pc                       & 12$^{\mathrm h} 30^{\mathrm m}29\fs5096$ & $+41\degr39\arcmin26\farcs886$ & 0.80  & 0.49 & 2.61 & 5.06 & 7.38 & 7.01 & 8.50 & 9.38 & 10.0 & 8.95 & 11.3 & 9.79 & 39.2 & 103  & 311  & 2237  & 1807  &  1.23E+40 & 8.34 \\
NGC 4490\tablenotemark{a} & 9.12 $\times$ 4.31 kpc, PA = 117$ ^\circ $ & 12$^{\mathrm h} 30^{\mathrm m}35\fs5393$ & $+41\degr38\arcmin42\farcs229$ & 48.6  & 79.4 & 172  & 401  & 543  & 633  & 719  & 784  & 930  & 757  & 572  & 499  & 1042 & 2374 & 4258 & 63119 & 90465 &  2.07E+41 & 10.1 \\
NGC 4485\tablenotemark{a} & 3.07 $\times$ 2.42 kpc,  PA = 20$ ^\circ $ & 12$^{\mathrm h} 30^{\mathrm m}30\fs6164$ & $+41\degr41\arcmin53\farcs385$ & 11.4  & 14.7 & 24.1 & 51.7 & 66.0 & 75.2 & 84.2 & 72.1 & 87.1 & 63.0 & 54.4 & 57.9 & 54.5 & 120  & 186  & 2689  & 5758  &  2.33E+40 & 8.86 \\
\enddata
\tablenotetext{a}{ North:  PA = 0$^\circ $, East: PA = 90$^\circ $ (see \autoref{fig:ARP269} for compass); PA refers to position of semi-major axis.}
\end{splitdeluxetable*}

\begin{deluxetable}{lcccccc}
\tablecaption{Bolometric Luminosity Measurements  \label{tab:lum}}
\tablewidth{0pt}
\tablehead{
\colhead{Object}   & \colhead{$\textit{L}_{160}$\tablenotemark{a}} & \colhead{$\textit{L}_{1100}$ MBB\tablenotemark{b}} & \colhead{$\textit{L}_{3}$\tablenotemark{c}} & \colhead{$\textit{L}_\mathrm{TIR}$\tablenotemark{d}} & \colhead{$\textit{L}_{1100}$ (3+TIR)\tablenotemark{e}} & \colhead{T$_{\text{d}}$\tablenotemark{f}} \\
 & \colhead{(\textit{L}$_\sun$)}                 &  \colhead{(\textit{L}$_\sun$)}                     &  \colhead{(\textit{L}$_\sun$)}                     &  \colhead{(\textit{L}$_\sun$)}  &  \colhead{(\textit{L}$_\sun$)}                         &  \colhead{(K)} } 
\startdata
Optical Nucleus (OPT) & 1.51E+09 & 	1.54E+09 & 	1.06E+09 & 	3.73E+08 & 	1.40E+09 & 	33.85 \\
Infrared Nucleus (IR) & 1.58E+09 & 	1.63E+09 & 	6.26E+08 & 7.63E+08 & 	1.39E+08 & 38.48 \\ 
\ion{H}{2} Region A	  & 9.72E+08 & 	1.00E+09 & 	3.69E+08 & 	5.55E+08 & 9.24E+08 & 	34.73 \\
\ion{H}{2} Region B	  & 6.94E+08 & 	7.26E+08 & 	1.66E+08 & 	4.74E+08 & 	6.39E+08 & 	35.45 \\ 
NGC 4490	          & 2.71E+10 & 	2.88E+10 & 	1.30E+10 & 	1.37E+10 & 	2.67E+10 & 	29.94 \\
NGC 4485	          & 2.35E+09 & 	2.47E+09 & 	1.65E+09 & 	7.11E+08 & 	2.36E+09 & 	27.01 \\
\enddata
\tablenotetext{a}{Integration between FUV and 160 $\mu$m.}
\tablenotetext{b}{Integration between FUV and 1100 $\mu$m. Single dust temperature (T$_{\text{d}}$), modified blackbody (MBB: $\nu^\beta \text{B}_{\nu}(\nu,\text{T}_\text{d})$, $\beta = 1.3$ from \citealt{Dun2000}) fit to the 70/160 flux density ratio, used for $\lambda > 160~\mu$m.}
\tablenotetext{c}{Integration between FUV and 3 $\mu$m (linear interpolation between 2.16 and 3.6 $\mu$m).}
\tablenotetext{d}{Empirical calibration from \citet{Dal2002}, where the 3 to 1100 $\mu$m luminosity (\textit{L}$_\mathrm{TIR}$) is estimated from the 24, 70, and 160 $\mu$m flux densities.}
\tablenotetext{e}{Sum of \textit{L}$_{\textbf{3}}$ and \textit{L}$_\mathrm{TIR}$.}
\tablenotetext{f}{Dust temperature for a $\beta=1.3$ modified blackbody (MBB) fit to the 70/160 flux density ratio.}
\end{deluxetable}

\subsection{Spectral Energy Distribution} \label{sec:sed}

To construct a spectral energy distribution (SED) for NGC 4485, NGC 4490, the infrared and optical nuclei, and the \ion{H}{2} regions A and B, we obtained photometry on sky-subtracted images from the FUV at 1539~\AA\ to the FIR at 160~$\mu$m using the apertures described in the previous section (see \autoref{tab:SEDMASSLUM}.  In all cases no local background subtraction was made. The photometry is summarized in \autoref{tab:SEDMASSLUM}, and the SEDs are shown in \autoref{fig:SEDGAL500}. In \autoref{fig:SEDGAL500} the SED has been extended to 1100 $\mu$m using a single dust temperature (T$_{\text{d}}$) modified blackbody, MBB: $\nu^\beta \text{B}_{\nu}(\nu,\text{T}_\text{d})$. For the dust emissivity we used $\beta = 1.3$ \citep{Dun2000}, and the dust temperature (T$_\text{d}$) was obtained by matching the (70/160) flux density ratio.  See column 7 of \autoref{tab:lum} for the dust temperatures (T$_{\text{d}}$).     We see that all six of the SEDs show a double-peak structure, as would be expected for a combination of stars and heated dust. Both of the galaxies have similar SEDs, although clearly NGC 4490 is much more luminous than NGC 4485. The shape of both galaxy SEDs is very similar to the shape of the infrared and optical nuclei SEDs. 

We measured the bolometric (FUV to 1100 $\mu$m) luminosity for the objects in two ways. In the first technique we extended the SED beyond 160~$\mu$m using a MBB; $\nu^\beta \text{B}_{\nu}(\nu,\text{T}_\text{d})$, with $\beta = 1.3$ and T$_\text{d}$ set by the (70/160) flux density ratio (see column 7 of \autoref{tab:lum}). The FUV to 1100 $\mu$m SED (see \autoref{fig:SEDGAL500}) was numerically integrated using the trapezoidal rule (i.e., using a piecewise linear approximation for the SED), and the resulting luminosity, $\textit{L}_{1100}~\mathrm{MBB}$, is shown in column 3 of \autoref{tab:lum}. For reference, we also show the luminosity obtained when the integration omits the FIR ($\lambda >$ 160 $ \mu $m) and covers the range 0.1539 $\leqslant \lambda \leqslant$ 160 $ \mu $m ($\textit{L}_{160}$ in column 2 of \autoref{tab:lum}).

In the second technique we first integrated between FUV and 3~$\mu$m (\textit{L}$_3$, column 4 in \autoref{tab:lum}), where the 3~$\mu$m point was estimated using a linear interpolation between the 2.16 and 3.6 $\mu$m flux densities. To account for emission beyond 3 $\mu$m we then used the empirical calibration from \citet{Dal2002}, where the 3 to 1100 $\mu$m luminosity (\textit{L}$_\mathrm{TIR}$) can be estimated from the 24, 70, and 160 $\mu$m flux densities. $\textit{L}_\mathrm{TIR}$ is shown in column 5 of \autoref{tab:lum} and the bolometric luminosity estimate, $\textit{L}_{1100}$ ({3+TIR), found by combining $\textit{L}_3$ and $\textit{L}_\mathrm{TIR}$ is shown in column 6. Inspection of \autoref{tab:lum} shows that the luminosity found in this manner is (5--15\%) lower than the luminosity found using the MBB. For simplicity we will adopt the MBB bolometric values in further discussions. 

We find that both the infrared and optical nuclei have similar bolometric luminosities: $L_\mathrm{bol}(\mathrm{OPT}) \approx 0.94 L_\mathrm{bol}(\mathrm{IR})$. They each contribute about five percent to the total luminosity of NGC 4490, and are each about 60 percent as luminous as the entire NGC 4485 galaxy. The \ion{H}{2} regions A and B both have a lower, but still significant, bolometric luminosity compared to either nuclei. For region A: $L_\mathrm{bol}(\mathrm{A}) \approx 0.65 L_\mathrm{bol}(\mathrm{OPT})$ and $L_\mathrm{bol}(\mathrm{A})\approx 0.61 L_\mathrm{bol}(\mathrm{IR})$. Similarly for region B: $L_\mathrm{bol}(\mathrm{B}) \approx 0.47 L_\mathrm{bol}(\mathrm{OPT})$ and $L_\mathrm{bol}(\mathrm{B})\approx 0.45 L_\mathrm{bol}(\mathrm{IR})$. We discuss the nucleus and \ion{H}{2} region luminosities further in \autoref{sec:sbt}.

\begin{figure}
\plotone{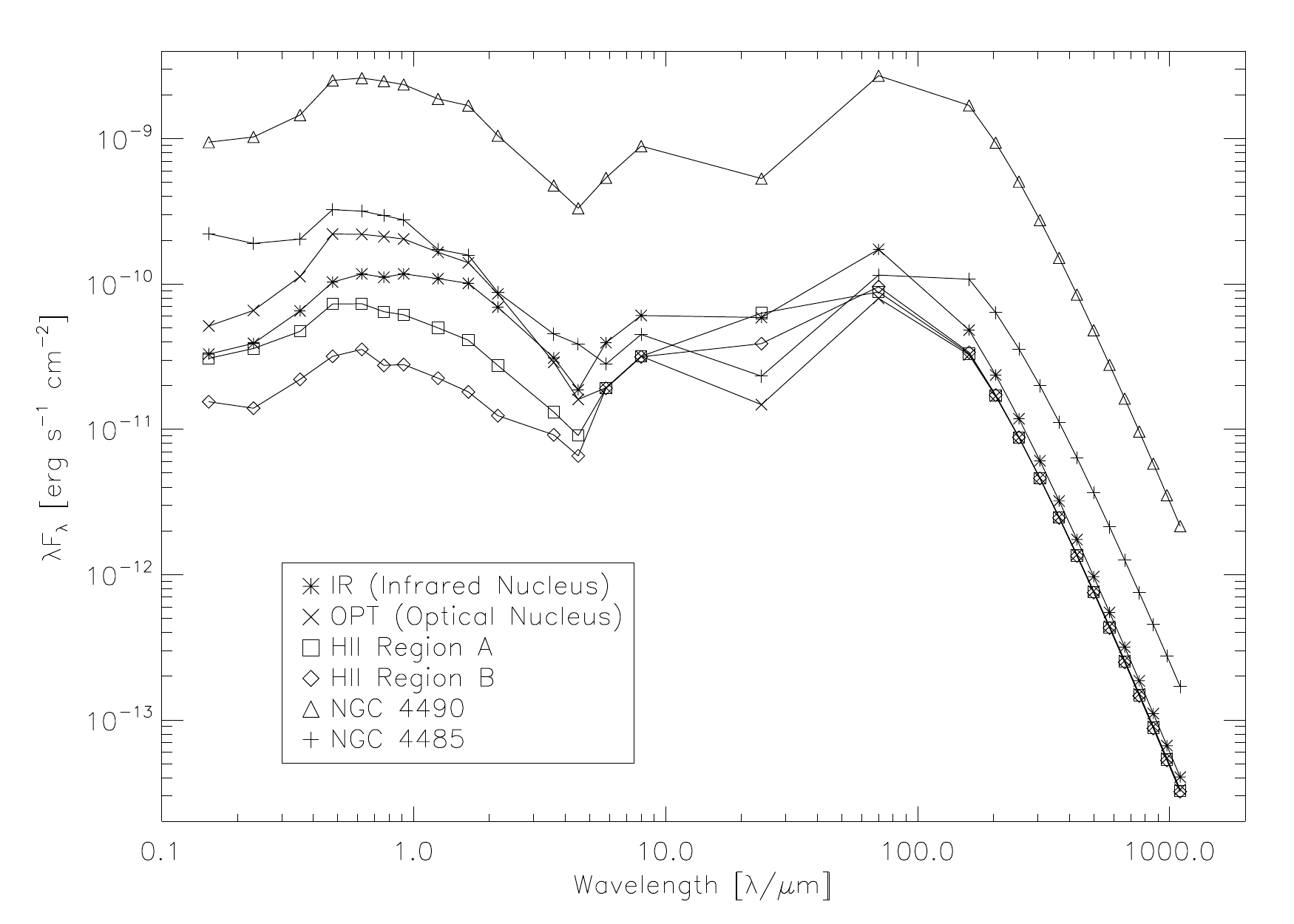}
\caption{Comparison of galaxy (NGC 4490 and NGC 4485) and clump SEDs. IR and OPT are the infrared and optical nuclei respectively of NGC 4490. A and B are \ion{H}{2} regions within NGC 4490. \autoref{fig:aps} shows the 500 pc radius aperture locations used for the clump photometry on a 3.6~$\mu$m image of NGC 4490. J2000 positions for the galaxy and clump apertures are given in \autoref{tab:SEDMASSLUM}.  \label{fig:SEDGAL500}} 
\end{figure} 

\subsection{Radial Profiles} \label{sec:rad}

We also compare the optical and infrared nuclei, and the two \ion{H}{2} regions A and B, by examining their 3.6 $\mu$m radial surface brightness profiles constructed using IDL. The same apertures used for photometry were divided into a series of concentric annuli of width 0.75 arcsec (3.6 $ \mu $m image pixel size, $ \approx $ 33.45 pc \textbf{at} 9.2 Mpc), and the average surface brightness within each annulus was calculated. The resulting normalized radial profiles are shown in \autoref{fig:RADIAL}.

\ion{H}{2} regions A and B both have sharply peaked profiles showing that the emission within the aperture is dominated by a single compact central source. In contrast, the OPT aperture profile shows a shallow, essentially linear, decrease in brightness. The IR aperture profile is a combination of these two profile types. There is a steep drop off at small distance, like the \ion{H}{2} region profiles, followed by a more shallow linear decrease similar to the OPT profile. There are a number of compact \ion{H}{2} regions within the IR aperture, as can be seen in the H$ \alpha $ image (\autoref{fig:HASOURCE}). In particular, \ion{H}{2} region 13 from \citet{Bou1970} is very close to the center of the aperture and is the cause of the central peak in the radial profile. Some smaller compact sources, not cataloged by \citet{Bou1970}, are responsible for the departure from linearity observed around a radius of 10 pixels.

In order to separate the compact and diffuse emission components in these regions we applied the spatial filtering technique described in \citet{Sof1979}. Starting with an initial image, denoting the surface brightness at any given position as $I$, a smoothed version of the image $I_\mathrm{s}$ is created using a Gaussian smoothing kernel. A new image, $I_\mathrm{clip}$, is calculated, where $I_\mathrm{clip} = I$ if $I-I_\mathrm{s} \leq 0$, and $I_\mathrm{clip} = I_\mathrm{s}$ if $I-I_\mathrm{s} > 0$. The $I_\mathrm{clip}$ image is then smoothed using the same Gaussian kernel creating a large-scale structure image, $I_\mathrm{large}$. A map of the small-scale emission structure is calculated as $I_\mathrm{small} = I - I_\mathrm{large}$. The exact spatial scale isolated using this technique is controlled by the size of the Gaussian kernel. For our data we used a Gaussian kernel with a FWHM $= 5.3$ arcsec, effectively removing point-source like emission from the $3.6~\mu$m image. The initial 3.6~$\mu$m image $I$ and the resulting $I_\mathrm{small}$ and $I_\mathrm{large}$ images are shown in the upper row and lower left panel of \autoref{fig:SMCOMP}. Comparison of the small-scale structure image with the H$\alpha$ image shown in the lower right panel of \autoref{fig:SMCOMP} shows the filtering technique has done a good job of isolating compact emission features likely associated with \ion{H}{2} regions.

Using the same technique described before, we obtained radial profiles for the optical and infrared nuclei as seen in the large-scale structure image (upper right panel of \autoref{fig:SMCOMP}). As expected, there is only a minimal change to the normalized profile of the optical nucleus. In contrast, the radial profile of the infrared nucleus now closely resembles that of the optical nucleus; there is a shallow linear decrease in intensity over the entire aperture. 

\begin{figure}
\plotone{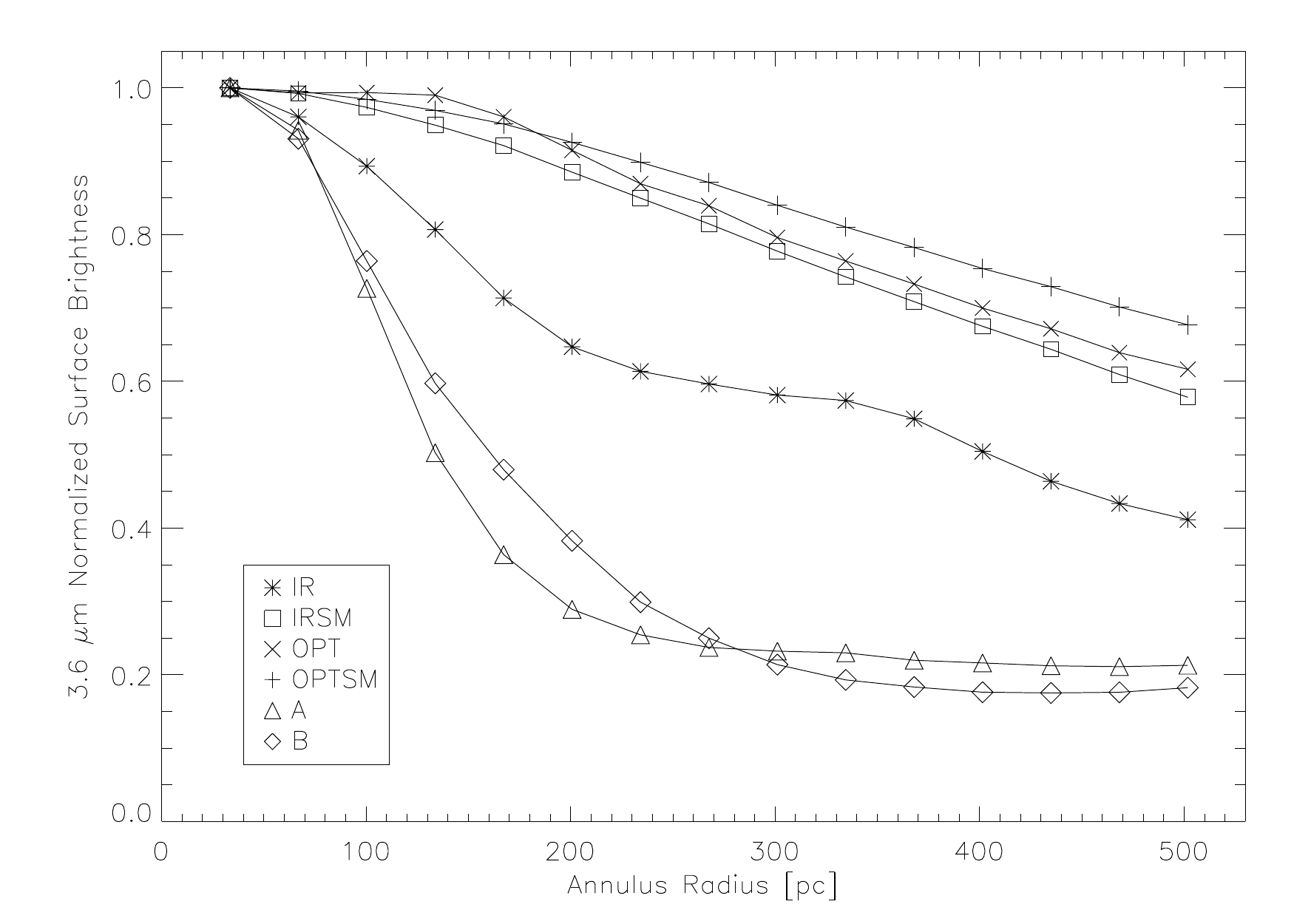}
\caption{3.6 $ \mu $m normalized surface brightness profiles. Each line shows the azimuthally-averaged normalized surface brightness profile for the 500 pc radius apertures surrounding the regions of interest in NGC 4490. OPT, IR, A, and B refer to the nuclei and \ion{H}{2} regions shown in \autoref{fig:aps}. OPTSM and IRSM refer to the profiles obtained using the OPT and IR apertures on the large-scale structure image (upper right panel of \autoref{fig:SMCOMP}).  \label{fig:RADIAL} }  
\end{figure}  

\begin{figure*}
\gridline{\fig{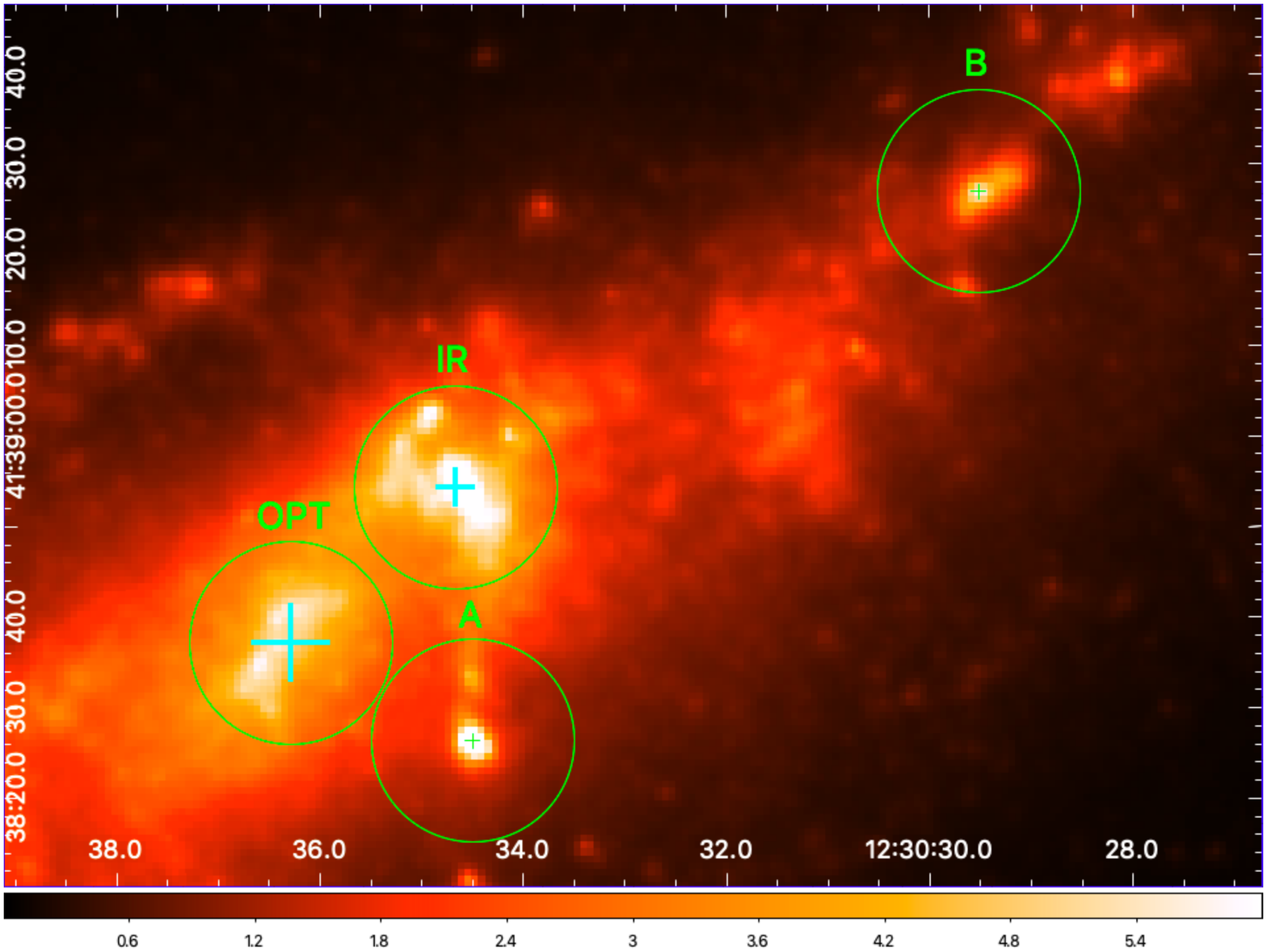}{0.5\textwidth}{3.6 $\mu$m -- Original}
\fig{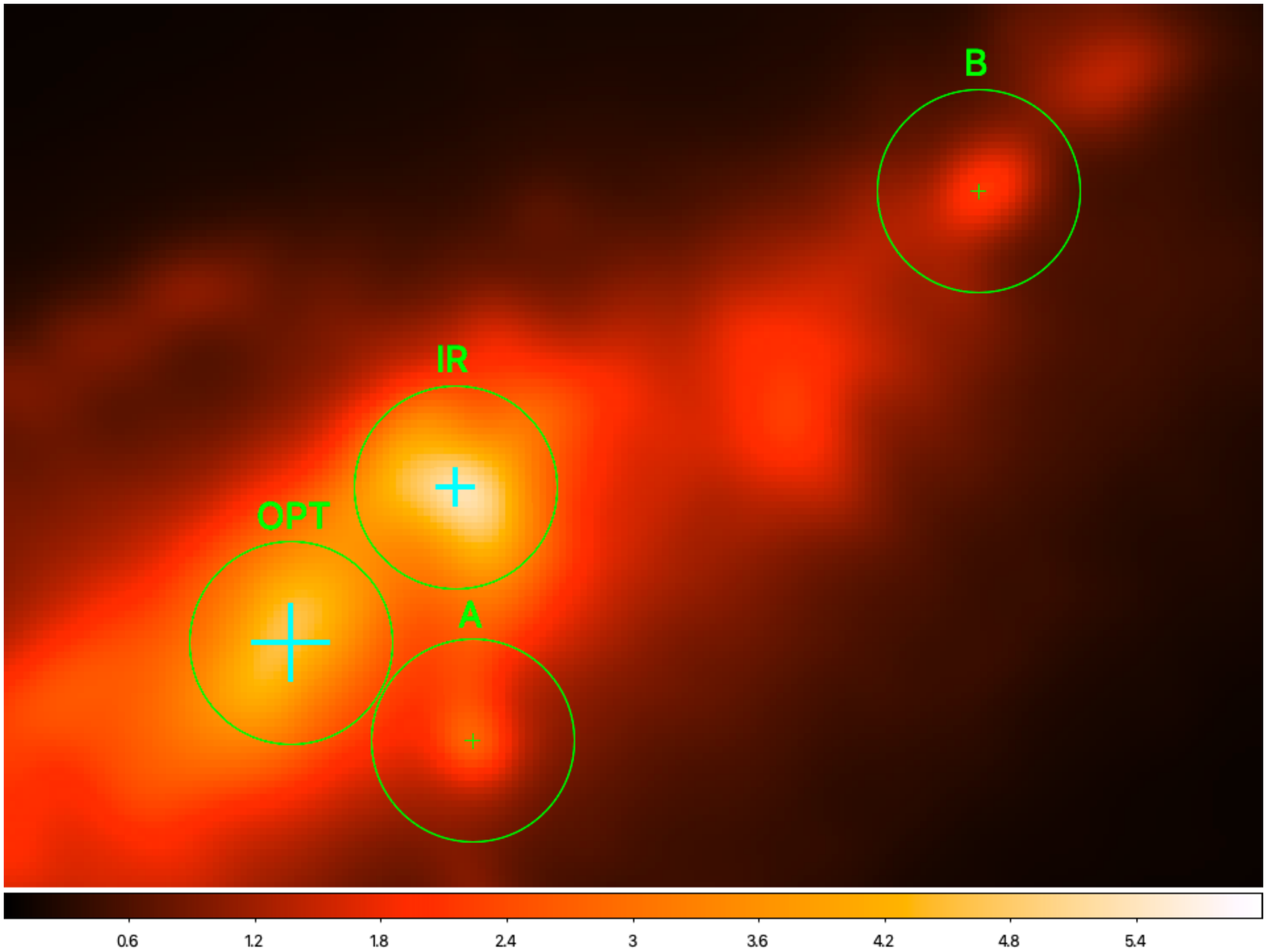}{0.5\textwidth}{3.6 $\mu$m -- Large Scale Structure }
          }
\gridline{\fig{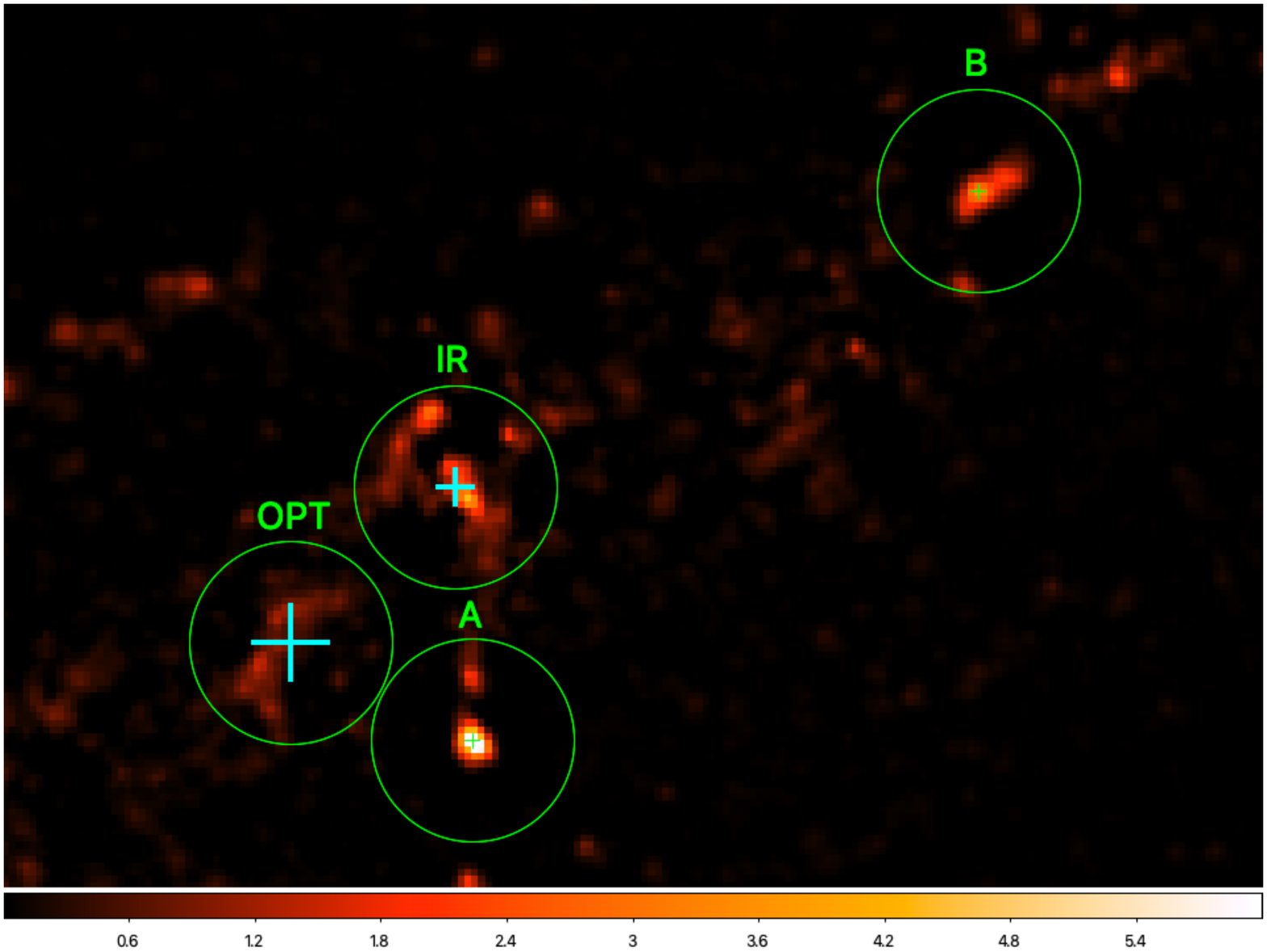}{0.5\textwidth}{3.6 $\mu$m -- Small Scale Structure}
\fig{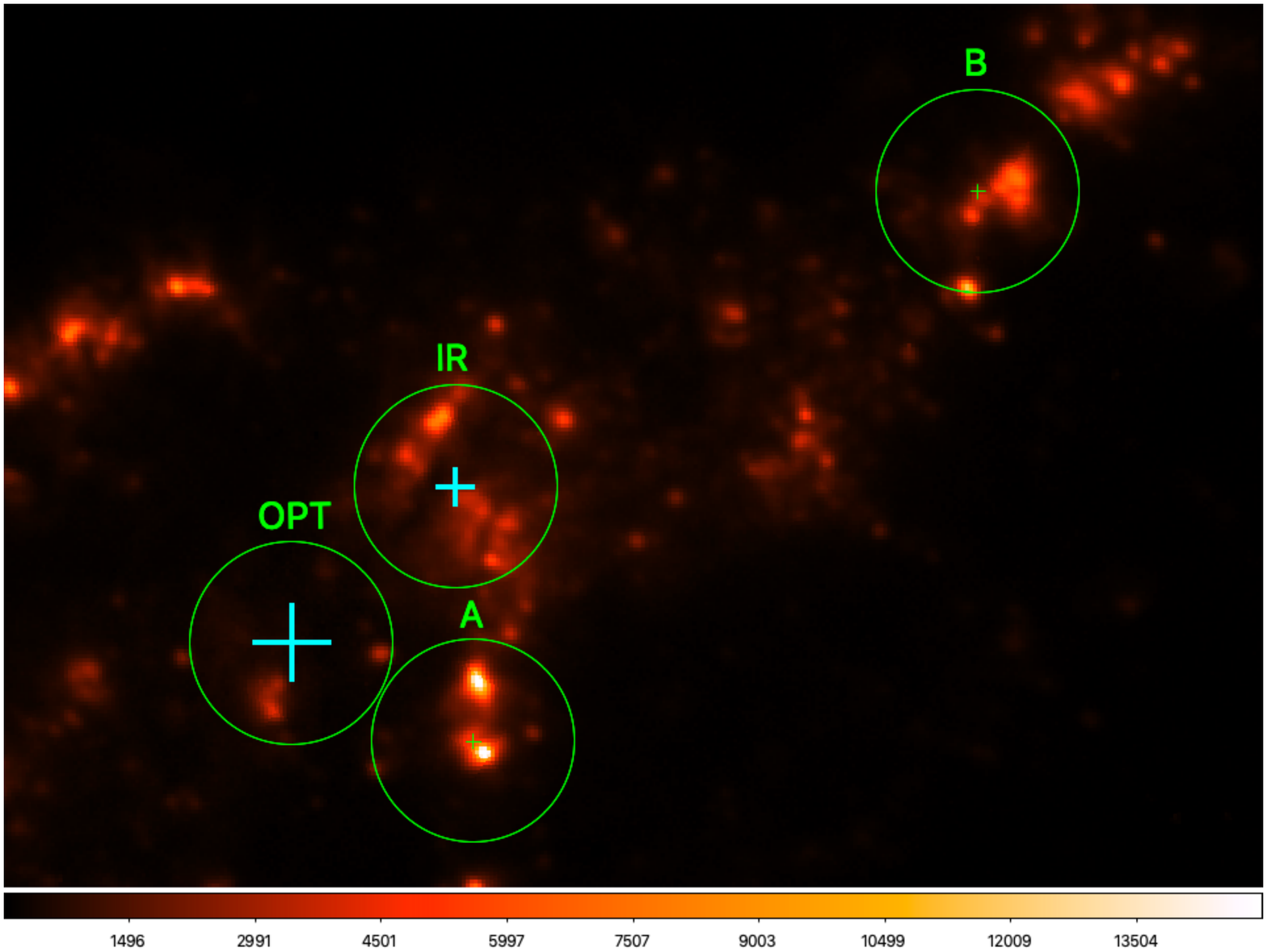}{0.5\textwidth}{H$ \alpha $}
          }
\caption{Spatial filtering of the NGC 4490 3.6 $\mu$m image ($\approx 6.2 \times 4.3$ kpc at 9.2 Mpc).  The large and small cyan crosses mark the locations of the optical and infrared nucleus respectively (see \autoref{fig:ARP269}) and the 500 pc radius apertures (green circles) are from \autoref{fig:aps}. The 3.6 $ \mu $m large scale structure was created using a Gaussian smoothing kernel with a FWHM of 5.3 arcsec.  Units are the same as in \autoref{fig:aps} and \autoref{fig:HASOURCE} for the 3.6 $\mu$m and H$\alpha$ images respectively.  Coordinate system is J2000. \label{fig:SMCOMP}}
\end{figure*}

\subsection{Star Formation Rate} \label{sec:sfr}

The combination of FUV and 24 $\mu$m IR emission has proven to be a good tracer of the star formation rate (SFR) of an extragalactic system \citep{Cal2007,Ler2008,Yim2016}. To calculate this quantity our \emph{GALEX} FUV and \emph{Spitzer} 24 $\mu$m images of NGC 4490/85 were first reprojected to a common pixel grid using \textsc{montage} \citep{Ber2017}, and the FUV image was converted from counts s$^{-1}$ to MJy sr$^{-1}$ to match the 24 $\mu$m data. The SFR surface density, $\Sigma_\mathrm{SFR}$, was then calculated on a pixel by pixel basis using the following equation from \citet{Ler2008}:
\begin{equation}
    \label{eqn:SFR} 
    \Sigma_\mathrm{SFR} = \left(8.1 \times 10^{-2} I_\mathrm{FUV} + 3.2^{+1.2}_{-0.7} \times 10^{-3} I_\mathrm{24}\right) \cos(i) \;\; \mathrm{M_\sun yr^{-1} kpc^{-2}}, 
\end{equation}
where $I_\mathrm{FUV}$ and $I_\mathrm{24}$ are the FUV and 24~$\mu$m surface brightness in MJy sr$^{-1}$, and $i$ is the inclination angle of the galaxy. $\Sigma_\mathrm{SFR}$ was then integrated over the appropriate aperture to obtain the total SFR ($\mathrm{M}_\sun$ yr$^{-1}$) for the region of interest. To facilitate comparison of our results with earlier estimates of the SFR, we multiplied our SFR values by a factor of 1.59 \citep{Ler2008}. This converts the SFR derived using \autoref{eqn:SFR}, which is based on a \cite{Kro2001} initial mass function (IMF) with a stellar mass range of $\mathrm{M}_{\star} $ = 0.1 -- 120 $ \mathrm{M}_{\odot} $, to a SFR appropriate for a \cite{Sal1955} IMF with a stellar mass range of $\mathrm{M} _{\star}$ = 0.1 -- 100 $ \mathrm{M}_{\odot}$. 

SFR was also calculated for the various regions with a combination of H$ \alpha $ and 24 $ \mu m$ using the following equation from \cite{Cal2007}:
\begin{equation}
    \label{eqn:SFRHA}
    \mathrm{SFR} = 5.3 \times 10^{-42}\left[ \textit{L}(\text{H}\alpha) + (0.031 \pm 0.006 \textit{L}(\mathrm{24 \,\mu m)}\right]  \mathrm{\ \  M_\sun yr^{-1}}, 
\end{equation}
where the luminosities are in erg s$^{-1}$ and $\textit{L}(\mathrm{24 \,\mu m)} $ is expressed as $ \nu $\textit{L}$_\nu$.  As before, we multiplied our SFR values by a factor of 1.59 to convert the SFR derived using \autoref{eqn:SFRHA}, which is which is based on a \cite{Kro2001} initial mass function (IMF) with a stellar mass range of $\mathrm{M}_{\star} $ = 0.1 -- 120 $ \mathrm{M}_{\odot} $, to a SFR appropriate for a \cite{Sal1955} IMF with a stellar mass range of $\mathrm{M} _{\star}$ = 0.1 -- 100 $\mathrm{M}_{\odot}$.

Our SFR calculations are summarized in the upper part of \autoref{tab:SFR}. Each region or galaxy is listed in column 1.  Column 4 gives the total SFR for each region calculated using \autoref{eqn:SFR} while Columns 2 and 3 of \autoref{tab:SFR} show the SFR that would be obtained using only the first or second terms of \autoref{eqn:SFR} respectively. The ratio of these quantities, which provides a measure of the degree to which the star formation activity is obscured or embedded, is shown in column 6 of \autoref{tab:SFR}. The SFR calculated using \autoref{eqn:SFRHA} is shown in column 5. In the lower part of \autoref{tab:SFR} we list two SFR estimates for NGC 4490 from the literature. \citet{Cle1999} obtained 4.7 $  \mathrm{M}_{\odot}$ yr$^{-1}$ as the SFR of NGC 4490 using radio emission and an extended Miller-Scalo IMF ($ \mathrm{M} _{\star} $ = 0.1 -- 100 $  \mathrm{M}_{\odot} $), which is similar to the Salpeter IMF \citep{Ken1983}.  \citet{Cle2002} used the H$\alpha$ luminosity and a Salpeter IMF ($ \mathrm{M} _{\star} $ = 0.1 -- 100 $  \mathrm{M}_{\odot} $) to obtain a SFR of 3.4 $  \mathrm{M}_{\odot} $ yr$^{-1}$ for NGC 4490.

The similarity of the SFR for the IR aperture and the \ion{H}{2} regions A and B is because there are compact \ion{H}{2} regions within this aperture as discussed in the previous section. Comparing columns 4 and 5, we see that the SFR estimates using the two different techniques agree to within the errors associated with the parameter uncertainties in each formula. Comparing the NGC 4490 results in columns 4 and 5 with the results shown in the lower part of the table we see that our H$\alpha$ value agrees with \citet{Cle2002}, and our FUV value is slightly lower. Unfortunately the \citet{Cle2002} and \citet{Cle1999} values do not have reported uncertainties, but we can conclude that all of the SFR values derived from the different techniques for NGC 4490 agree to within a factor of two or better.

\begin{deluxetable*}{cccccc}
\tablecaption{Star Formation Rate (SFR) Calculations  \label{tab:SFR}}
\tablewidth{0pt}
\tablehead{
\colhead{Object}   & \colhead{SFR$ _{FUV} $} & \colhead{SFR$ _{24\;\mu m} $} &\colhead{SFR$ _{FUV+24\mu m} $\tablenotemark{a}}  & \colhead{SFR$ _{H\alpha+24\mu m} $\tablenotemark{b}} & \colhead{SFR$_{24\;\mu m/FUV}$}\\  & \colhead{[$ \mathrm{M}_{\odot} $/yr]} & \colhead{[$ \mathrm{M}_{\odot} $/yr]} & \colhead{[$ \mathrm{M}_{\odot} $/yr]} & \colhead{[$ \mathrm{M}_{\odot} $/yr]}& \colhead{Ratio}}
\startdata
Optical Nucleus (OPT) & 0.029 & 0.054$ _{\,0.042}^{\,0.074} $ & 0.084$ _{\,0.072}^{\,0.10} $ & 0.092$ _{\,.085}^{\,0.10} $ & 1.8$ _{\,1.4}^{\,2.5} $\\
Infrared Nucleus (IR)   & 0.018 & 0.13$ _{\,0.10}^{\,0.17} $ & 0.14$ _{\,0.11}^{\,0.19} $ & 0.32$ _{\,0.29} ^{\,0.35}$ & 12$ _{\,9}^{\,16} $\\
\ion{H}{2} Region A     & 0.019 & 0.22$ _{\,0.17}^{\,0.31} $ & 0.24$ _{\,0.19}^{\,0.33} $ & 0.29$ _{\,0.26}^{\,0.32} $ & 12$ _{\,9}^{\,16} $\\
\ion{H}{2} Region B  & 0.0084 & 0.14$ _{\,0.11}^{\,0.19} $ & 0.15$ _{\,0.12}^{\,0.20} $ & 0.21$ _{\,0.19}^{\,0.23} $ & 16$ _{\,13} ^{\,22}$\\
NGC 4485  & 0.12 & 0.080$ _{\,0.062}^{\,0.11} $ & 0.20$ _{\,0.19}^{\,0.23} $ & 0.26$ _{\,0.25}^{\,0.27} $  & 0.65$ _{\,0.50} ^{\,0.88}$ \\
NGC 4490   & 0.53 & 1.8$ _{\,1.4}^{\,2.5} $  & 2.4$ _{\,2.0} ^{\,3.1}$  & 3.15$_{ \,2.88}^{ \,3.42}  $ & 3.4$ _{\,2.7}^{\,4.7} $\\
\hline \ 
NGC 4490\tablenotemark{c} & & &  3.4\\
NGC 4490\tablenotemark{d} & & &  4.7\\
\enddata
\tablecomments{Superscript and subscript values represent the range of calculated values consistent with the parameter uncertainties in the applicable SFR formula.}
\tablenotetext{a}{SFR calculation procedure from \cite{Ler2008} with Salpeter IMF ($ \mathrm{M} _{\star} $ = 0.1 -- 100 $  \mathrm{M}_{\odot} $) using FUV and $24\;\mu$m data.}
\tablenotetext{b}{SFR calculation procedure from \cite{Cal2007} with Salpeter IMF ($ \mathrm{M} _{\star} $ = 0.1 -- 100 $  \mathrm{M}_{\odot} $) using H$ \alpha $ and $24\;\mu $m data.}
\tablenotetext{c}{From \citet{Cle2002}. SFR calculated using H$ \alpha $ luminosity with Salpeter IMF ($ \mathrm{M} _{\star} $ = 0.1 -- 100 $  \mathrm{M}_{\odot} $).}
\tablenotetext{d}{From \citet{Cle1999}. SFR calculated using integrated thermal emission from radio observations with an extended Miller-Scalo IMF ($ \mathrm{M} _{\star} $ = 0.1 -- 100 $  \mathrm{M}_{\odot} $).}
\end{deluxetable*}

\section{Discussion} \label{sec:dis}

 Clearly the infrared nucleus within NGC 4490 is a significant structural component of the galaxy. Our analysis has shown that, compared to the optical nucleus, it has a similar spatial extent, stellar mass, bolometric luminosity, and radial brightness profile at 3.6 $\mu$m. The feature is particularly intriguing as a double nucleus morphology is not a common structure seen in low redshift spiral galaxies. For example, the \emph{Spitzer} Infrared Nearby Galaxy Survey \citep[SINGS;][]{Ken2003} legacy project examined a sample of 75 nearby ($d < 30$~Mpc) galaxies representing the full range of Hubble types. While this is not a complete survey, it is interesting to note that none of the spiral galaxies observed as part of SINGS show a double-nucleus morphology in the near and mid infrared like we see in NGC 4490. In the following subsections we discuss possible origins of the infrared nucleus/region and the different implications for the history of the Arp 269 system. We then compare our results to other observations of nuclei and star-formation regions found in interacting galaxies.
 
\subsection{A Merger Remnant?} \label{sec:merger}

Given the similarities between the IR feature and the optical nucleus, one natural interpretation is that the IR feature is the nucleus of a late-stage minor-merger remnant. Closely spaced double nuclei observed in other galaxies are typically interpreted as being due to a minor merger. For example, \citet{Gim2004} examined various major galaxy catalogs \citep[e.g.,][]{Pet1978, Maz1991} as well as the literature to identify spiral galaxies with reported double nuclei that are candidates for minor mergers. Their final catalog contains 107 spiral galaxies with $cz < 15 000$ km s$^{-1}$ or $\mathrm{m_B} < 18$, corresponding to distances $d < 200$ Mpc. We note that only two of the galaxies listed in this combined catalog have distances smaller than the distance to NGC 4490.

Since the pioneering study of \citet{Lar1978}, numerous studies have concluded that mergers/interactions between galaxies can trigger/create higher rates of star formation than in isolated galaxies (e.g., see the \citet{Smi2016} comparative study of 46 nearby ($d < 150$ Mpc) interacting galaxies). The connection between mergers and intense star formation is dramatically illustrated by the more extreme ultra-luminous infrared galaxies (ULIRGs). ULIRGs, defined by $\mathrm{L_{IR}} > 10^{12}$ L$_{\sun}$, are known to contain starbursts driven by interacting pairs of equal mass disk galaxies that are in advanced phases of merging (see \citealt{Kim2002}, \citealt{Vei2002}, and \citealt{Vei2006}). A thorough discussion of all aspects of galaxy collisions/interactions, including the association of star formation and collisions, can be found in the reviews by \citet{Bar1992}, \citet{Str1999}, and \citet{Str2006}.

In this picture the Arp 269 system should be viewed as a triple system, where NGC 4490 is a late-stage merger that is now interacting with NGC 4485. This earlier merger provides a possible driver for the NGC 4490 starburst. \citet{Pri2017} suggest that large-scale, extended star formation is  expected for interactions between low-mass galaxies, in contrast to higher-mass galaxy interactions which lead to a nuclear starburst.  \citet{Lot2008} examined merger time scales applicable to mergers between equal-size spiral galaxies, and found the full merger takes place on 1-3 Gyr time scales. The models of \citet{Pea2018} suggest that a similar timescale is also applicable for minor mergers.  The \citet{Lot2008} models also found that the star formation activity induced by the merger lasts for $\approx 10^8$ yr. These timescales compare favorably with the SF driven emplacement timescale of the NGC 4490 \ion{H}{1} plume of $6 \times 10^8$ yr \citep{Cle1998} and the estimated duration of continuous high levels of star formation in NGC 4490 (constant for $10^8$ years) from \citet{Cle1999}. 

Alternatively NGC 4490 may be a merger remnant, but the extended \ion{H}{1} plume may instead be primarily a tidal feature related to the merger of two gas-rich dwarf galaxies. The \citet{Pea2016} study of \ion{H}{1} in dwarf galaxy pairs found that a significant fraction of the \ion{H}{1} lies outside the optical extent of such galaxies so mergers involving dwarf pairs have the potential to create extensive \ion{H}{1} features. Using the \ion{H}{1} Rogues Gallery \citep{hib2001} as a guide, we found four comparable examples of gas-rich merging systems in the literature.  I Zw18 is a nearby ($\approx 10$ Mpc) blue compact dwarf galaxy embedded in a huge \ion{H}{1} envelope, and \citet{Zee1998a} consider a merger is one possible origin for the \ion{H}{1} morphology. A better analog may be the blue compact dwarf  II Zw40, which has a strikingly elongated \ion{H}{1} morphology. \citet{Zee1998b} consider II Zw40 to be a clear example of an ongoing merger between two gas-rich dwarf galaxies. NGC~2146 is a larger galaxy, but it also has highly elongated \ion{H}{1} filaments. \cite{Tar2001} suggest that these formed due to the interaction between NGC~2146 and a (no longer visible) low-surface brightness late-type gas-rich galaxy. Finally, \citet{Iyer2004} examined the Arp 158 system and showed that the \ion{H}{1} extends well beyond the optical extent of the system. They conclude that it is a mid-stage merger similar to the prototypical mid-stage merger system, Arp 157, discussed in \citet{Hib1996}. Each of these systems are possible analogs of NGC 4490, minus the additional interaction with NGC 4485.
 
A variation on this scenario is that the \ion{H}{1} surrounding Arp 269 is actually a large, highly warped, \ion{H}{1} disk. This picture was considered by \citet{Cle1998}, but they rejected it because there was no obvious way to obtain the degree of warping required. Given the similarity of the  mass and size of the optical and infrared nuclei a major-merger cannot be discounted, and could induce the necessary severe disk warping.  In this case, the \ion{H}{1} ``plume" would be the highly warped remnant of the progenitor disk(s).

\subsection{An Induced Star-Formation Complex?} \label{sec:sfcomplex}

In the \citet{Pea2018} model of the Arp 269 system, the need for a star-formation driven wind is removed as the \ion{H}{1} plume is primarily a tidal feature from the initial encounter between NGC 4485 and NGC 4490. In this scenario the feature we are calling the infrared nucleus would be an large region of star-formation activity induced by one or both or the NGC 4490/85 encounters. The star-formation activity associated with the clump may still contribute material to the \ion{H}{1} plume, but it is not the primary source of the plume.

In favor of this picture is the fact that the infrared nucleus region clearly contains signposts of active star formation activity (see \autoref{sec:rad}). In addition, radio continuum emission from the region corresponding to the infrared nucleus was first observed at 2.7 and 8.1 GHz by \citet{Sea1978}. These aperture synthesis observations had high enough resolution (9 and 3 arcsec respectively) to show clearly that the strongest radio emission from NGC 4490 is an extended non-thermal source located within an \ion{H}{2} region complex offset from the optical center of the galaxy. \citet{Sea1978} also show that the thermal radio emission from the \ion{H}{2} regions themselves contribute only a small amount to the total radio emission, and they suggest that the non-thermal emission arises from multiple supernova events. 

\subsection{Comparison to Clumps in Other Interacting Galaxies} \label{sec:sbt}

To distinguish between the above two possibilities for the nature of the infrared feature in NGC 4490, we compared the 3.6 $\mu$m luminosities and the derived stellar masses of the optical nucleus and the infrared nucleus/region in NGC 4490 with both nuclei and extra-nuclear star-forming regions within other interacting galaxies. As a comparison sample, we used the Spirals, Bridges, and Tails (SB\&T) sample of nearby pre-merger galaxy pairs studied by \citet{Smi2007,Smi2010,Smi2016}. Using 8.0 $\mu$m \emph{Spitzer} images, \citet{Smi2016} selected discrete regions of emission (``clumps"), which were then classified as disk clumps (likely star-forming regions/complexes) or nuclear clumps (likely galactic nuclei) based on their position and relative brightness in \emph{Spitzer} 3.6 $\mu$m images.

 \citet{Smi2016} did their clump photometry using apertures of 2.5 kpc and 1.0 kpc radii. To match our measurements for the nuclei in Arp 269, we re-did the \citet{Smi2016} clump and nuclear photometry on sky-subtracted images using smaller aperture radii of 500 pc and did not subtract a local background on the galaxy. We made aperture corrections to the photometry interpolating between the suggested values from the IRAC Instrument Handbook \footnote{http://irsa.ipac.caltech.edu/data/SPITZER/docs}. We limited our sample to the 18 SB\&T pairs with distances less than 43 Mpc. At that distance, a 500 pc radius corresponds to 2.4 arcsec, which is the minimum radius suggested for aperture corrections for IRAC. 
 
 In the upper and bottom left panels of \autoref{fig:SBTCLPCOMP}, we provide histograms of the 3.6 $\mu$m absolute magnitudes of the nuclei and the disk clumps, respectively. The red hashed areas give the location of the absolute 3.6 $\mu$m magnitudes of the optical nucleus and the infrared nucleus/region in NGC 4490.  For the disk and nuclear regions in the SB\&T sample, we calculated stellar masses from the 3.6 $\mu$m and 4.5 $\mu$m flux densities using \autoref{equ:STARMASS} (see \autoref{sec:STARMASS}). In the right panels of \autoref{fig:SBTCLPCOMP}, we compare the masses of the SB\&T nuclei (top panel) and disk clumps (bottom panel) with the optical nucleus and the infrared nucleus/region in NGC 4490.  This comparison shows that the OPT and IR regions in NGC 4490 are more luminous and more massive than almost all of the SB\&T disk clumps. If the possible second nucleus (IR) in NGC 4490 is only an extra-nuclear star-forming region, then it is both extremely luminous and very massive for such a region. In contrast, when we consider the SB\&T nuclear clumps, both of the NGC 4490 regions lie just slightly below the median 3.6 $\mu$m absolute magnitude and $\log(\mathrm{M}_{\star}$) bins.  In \autoref{fig:SBTCLPCOMP} we also show the values for the \ion{H}{2} regions A and B as blue hashed areas. We see that these regions fall well within the disk clump distributions and are located well below the median magnitude and mass values for nuclear clumps. 

Some idea of how these clumps compare to nuclei and \ion{H}{2} regions in a sample of (primarily) non-interacting galaxies can be obtained from data from \citet{Ken1989a}. In their Figure 2 they show the distribution of H$\alpha$ luminosities for a sample of approximately 80 nuclei (``\ion{H}{2} nuclei") and disk \ion{H}{2} regions. Comparing our measured H$\alpha$ luminosities with these distributions, we find that both the OPT and IR regions have H$\alpha$ luminosities ($\log\left(L_{\text{H}\alpha}\right) = 39.81$ and 40.29) that are higher than the average nuclei luminosity ($\log(L_{\text{H}\alpha} = 39.52$), but fall to the lower end of the starburst nuclei distribution. The A and B \ion{H}{2} regions, with $\log(L_{\text{H}\alpha}) = 40.16$ and 40.09) fall in the very high luminosity tail of the disk \ion{H}{2} regions. These results are similar to what we see with the SB\&T sample.

In general, the more luminous a galaxy, the more luminous the central nucleus/bulge. We illustrate this trend in \autoref{fig:SBTGALCOMP}, where we compare the 3.6 $\mu$m absolute magnitudes of the SB\&T nuclei with the total 3.6 $\mu$m absolute magnitudes of their host galaxy. A clear correlation is present and the NGC 4490 nuclei both fit this correlation.

In \autoref{fig:SFRSBT} we show the SFR for the disk clumps in the SB\&T sample along with the position of the \ion{H}{2} regions A and B. We see that SFR of these regions is on the high-end tail of the SFR distribution.

 \begin{figure}
\plotone{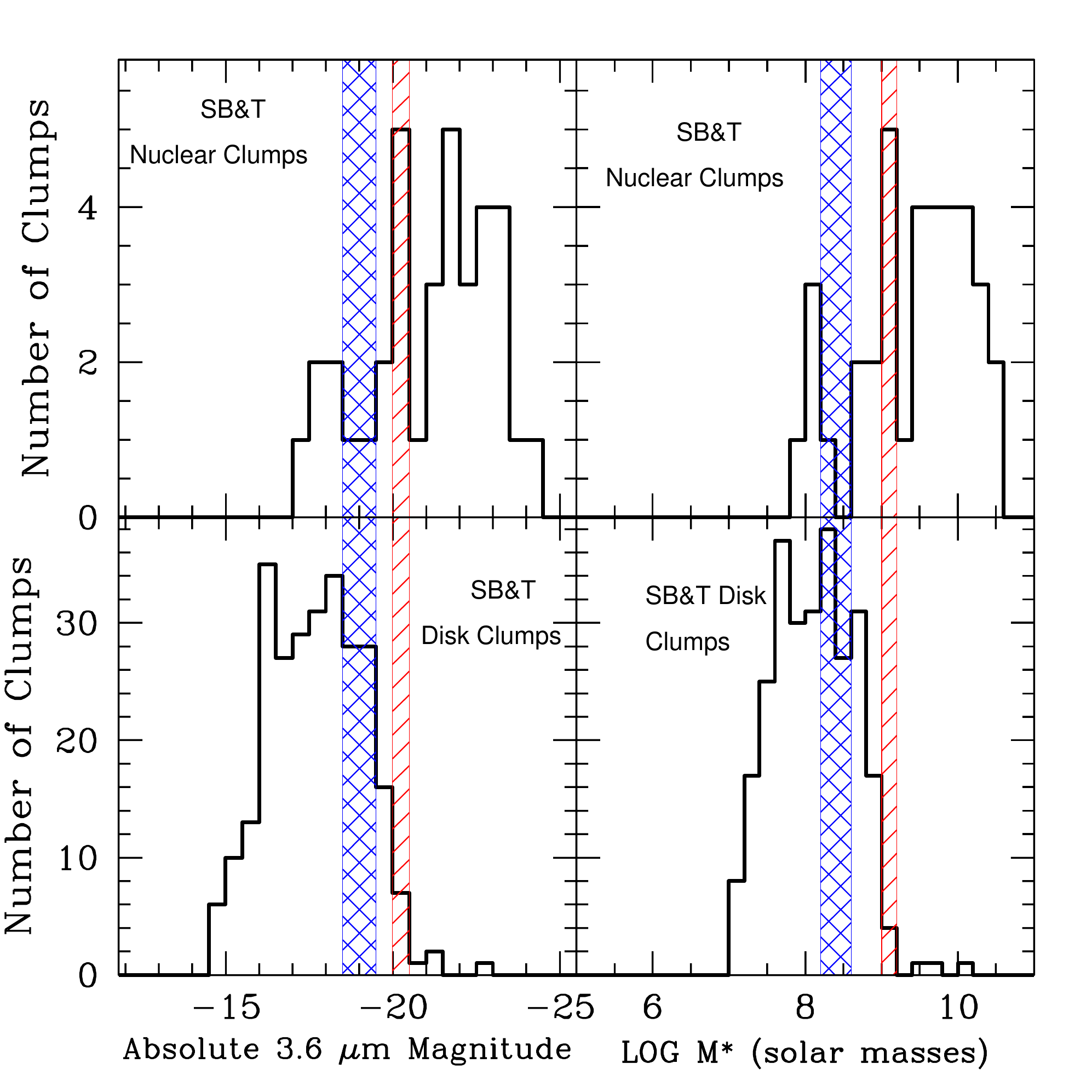}
\caption{SB\&T Comparison. Left column shows histograms of the clump luminosity at 3.6 $\mu$m. Right column shows histograms of the clump stellar mass derived from 3.6 and 4.5 $\mu$m photometry. The top row shows clumps that are likely galactic nuclei,  and the bottom row shows clumps that are likely star-forming regions/complexes. The red hashed areas shows our NGC 4490 results for both OPT and IR nucleus (the two nuclei have similar values), and the blue hashed areas show the A \& B clumps for the  \ion{H}{2} regions for NGC 4490.  The SB\&T photometry was redone using 500 pc radius apertures, and NGC 4490, which was in the original SB\&T sample, is not included in the SB\&T histograms. \label{fig:SBTCLPCOMP} }      
\end{figure}  

\begin{figure}
\plotone{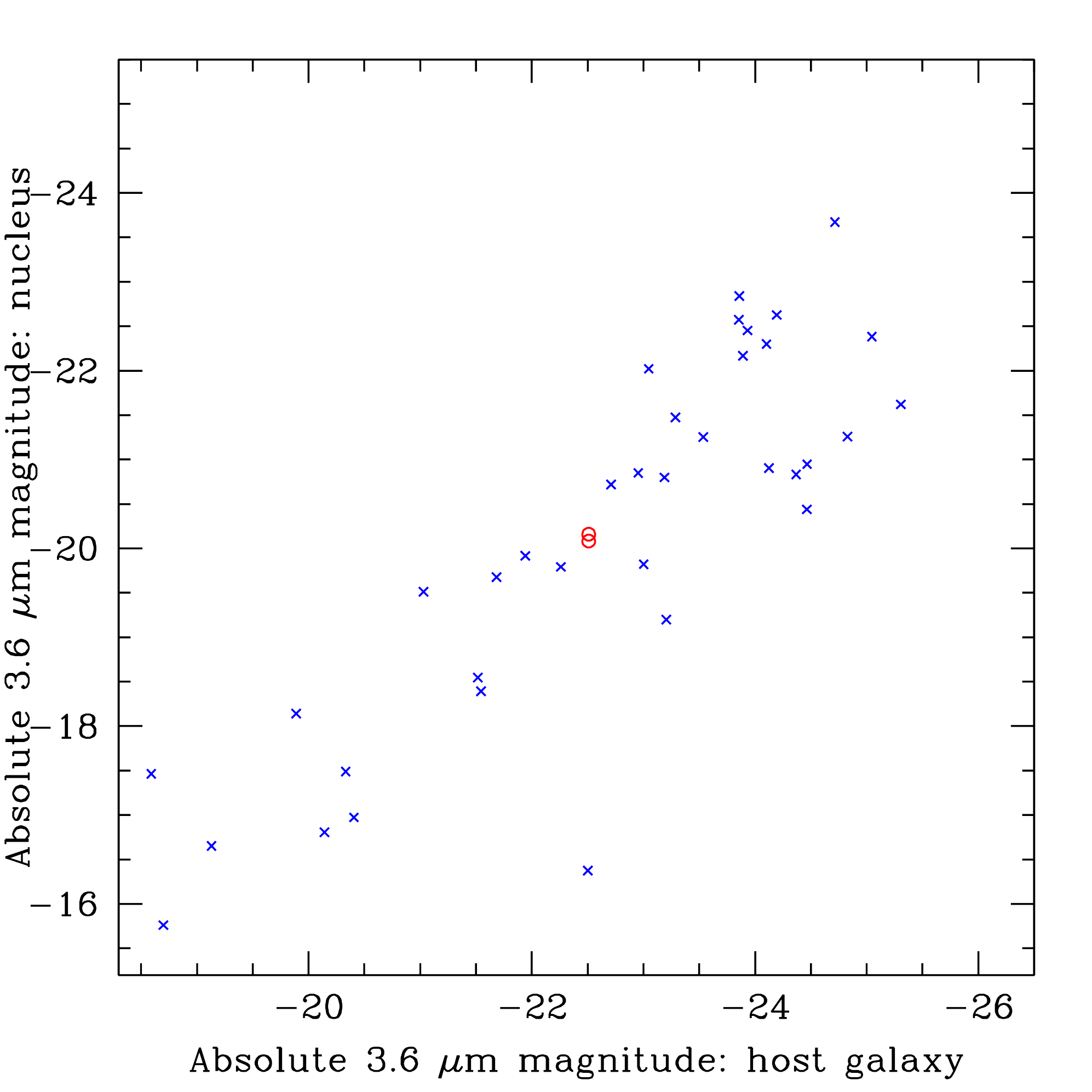}
\caption{Luminosity trend of nuclei and galaxy luminosity. The blue crosses show the 3.6 $\mu$m absolute magnitudes of the galactic nuclei in the SB\&T galaxy sample, plotted against the global 3.6 $\mu$m absolute magnitude for the host galaxy. The optical nucleus and the proposed infrared nucleus in NGC 4490 are plotted as open red circles.   \label{fig:SBTGALCOMP} }          
\end{figure}  

 \begin{figure}
\plotone{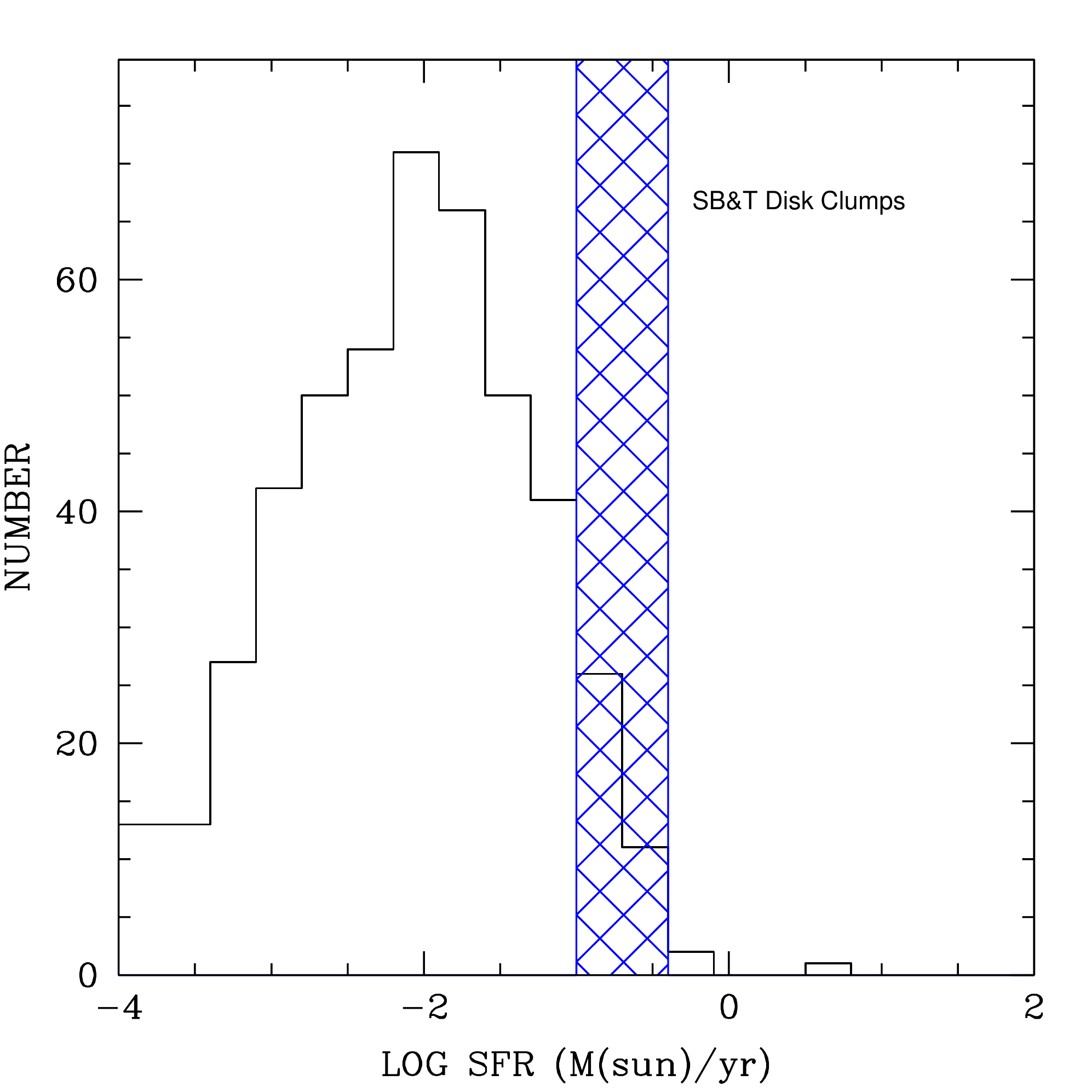}
\caption{SB\&T SFR Comparison. The histogram shows the distribution of the SFR calculated for SB\&T disk clumps. The blue hash region indicates the two bins where the NGC 4490 \ion{H}{2} region A \& B clumps would be located.  SFR calculation procedure from \citet{Ler2008,Liu2011} with Salpeter IMF ($\mathrm{M} _{\star} $ = 0.1 -- 100 $ \mathrm{M}_{\odot} $), using FUV and $24\;\mu $m.  \label{fig:SFRSBT} }  
\end{figure}

\section{Conclusions} \label{sec:con}

NGC 4490 has a clear double nucleus structure. One nuclei is optically visible while the other is dust shrouded and is prominent at infrared and radio wavelengths. The double nucleus morphology is best observed at NIR wavelengths.

The infrared nucleus has similar physical properties as the optical nucleus. Comparison with a sample of SB\&T pre-merger galaxy pairs show that the luminosity and stellar mass of the feature is similar to other galactic nuclei and much larger than most extra-nuclear star-forming regions. The most straightforward interpretation of the observations is that NGC 4490 is itself a late-stage merger remnant. Such a merger could drive the extended high level of star formation activity required for a star-formation driven emplacement of the Arp 269 \ion{H}{1} plume. Variations on this merger scenario would have the extended \ion{H}{1} morphology arising either as a tidal feature related to the merger, or as a highly warped disk. An alternative interpretation is that the infrared nucleus is actually a very large active region of star formation induced by the encounter(s) between NGC 4490 and NGC 4485. In this picture the \ion{H}{1} plume is primarily a tidal feature, but the intense star-formation activity associated with the infrared region would be expected to contribute material to the plume via stellar and supernova driven outflows.

Regardless of the merger history of the Arp 269 system, and the exact emplacement mechanism of the \ion{H}{1} plume (tidal, disk, non-tidal, combination), we have shown that the infrared nucleus/feature in NGC 4490 is a significant structure. Its origin and influence should be considered in any future studies of the Arp 269 system.

\acknowledgments

This research has made use of the NASA/IPAC Extragalactic Database (NED) which is operated by the Jet Propulsion Laboratory, California Institute of Technology, under contract with the National Aeronautics and Space Administration. This work is based in part on observations made with the \emph{Spitzer} Space Telescope, which is operated by the Jet Propulsion Laboratory, California Institute of Technology under a contract with NASA.  This research has made use of \textit{GALEX} products.  \textit{GALEX} was a NASA mission managed by the Jet Propulsion Laboratory. The GALEX science team was based at the California Institute of Technology.  Funding for SDSS-III has been provided by the Alfred P. Sloan Foundation, the Participating Institutions, the National Science Foundation, and the U.S. Department of Energy Office of Science.  SDSS-III is managed by the Astrophysical Research Consortium for the Participating Institutions of the SDSS-III Collaboration.  This research made use of \textsc{montage} \citep{Ber2017}. The development of \textsc{montage} is funded by the National Science Foundation under Grant Number ACI-1440620, and was previously funded by the National Aeronautics and Space Administration's Earth Science Technology Office, Computation Technologies Project, under Cooperative Agreement Number NCC5-626 between NASA and the California Institute of Technology.   The Westerbork Synthesis Radio Telescope (WSRT) is operated by ASTRON (Netherlands Foundation for Research in Astronomy) with support from The Netherlands Foundation for Scientific Research (NWO).  An early stage of this research made use of data products from the Wide-field Infrared Survey Explorer (\textit{WISE}), which is a joint project of the University of California, Los Angeles, and the Jet Propulsion Laboratory/California Institute of Technology, funded by the National Aeronautics and Space Administration.   The authors wish to thank Jay Gallagher of the University of Wisconsin for initial help with the research.  B.J.S. acknowledges support from National Science Foundation Extragalactic Astronomy grant ASTR-1714491.

\vspace{5mm}
\facilities{Bok, \textit{GALEX}, FLWO:2MASS, NED, Sloan (SDSS), \textit{Spitzer}, VLA, \textit{WISE}, WSRT}

\software{IDL,   GALFIT \citep{Pen2002,Pen2010},  MONTAGE \citep{Ber2017}}

\bibliography{ref}{}
\bibliographystyle{aasjournal}

\end{document}